\documentstyle[12pt,a4,epsfig]{article}

\newcommand{\beq}{\begin{equation}}
\newcommand{\eeq}{\end{equation}}
\newcommand{\beqar}{\begin{eqnarray}}
\newcommand{\eeqar}{\end{eqnarray}}
\newcommand{\PR}{Phys. Rev.\ }

\newcommand{\PL}{Phys. Lett.\ }
 \newcommand{\NP}{Nucl. Phys.\ }

\begin{document}
\vspace*{-2cm}
\hfill NTZ 26/2000

\vspace{4cm}
\begin{center}
{\LARGE \bf Parton interactions \\ in the Bjorken limit of QCD
\footnote{Supported by Deutsche
Forschungsgemeinschaft KI 623/1-2, \ \ based on our contribution
to the workshop in honour of Lev N. Lipatov's 60th birthday        
}}\\[2mm]
\vspace{1cm}
S.E. Derkachov$^{\dagger \#}$ and R.~Kirschner$^\dagger$
 \vspace{1cm}

$^\dagger$Naturwissenschaftlich-Theoretisches Zentrum  \\
und Institut f\"ur Theoretische Physik, Universit\"at Leipzig
\\
Augustusplatz 10, D-04109 Leipzig, Germany
\\ \vspace{2em}
$^{\#}$
Department of Mathematics, St. Petersburg Technology Institute, \\
St. Petersburg, Russia
\end{center}

\vspace{1cm}
\noindent{\bf Abstract:}

We consider the Bjorken limit in the framework of the effective action
approach and discuss its similarities to the Regge limit. The proposed
effective action allows for a rather simple calculation of the known
evolution kernels. We represent the result in terms of two-parton
interaction operators involving gluon and quark operators depending on
light-ray position and helicity and analyze their symmetry properties.

\vspace{3cm}

\vspace*{\fill}
\eject
\newpage

\section{Introduction}
\setcounter{equation}{0}

During three decades high energy  processes related to the
Bjorken limit of scattering amplitudes are playing a major role in the
investigation of the hadronic structure and the interaction of hadronic
constituents. Quite a number of theoretical concepts and methods have
been developed resulting by now in a standard treatment described in
textbooks \cite{JCPS} and reviews \cite{GStBLKod}.

In the last years increasing attention is devoted to topics going beyond
the standard deep inelastic structure functions, among them the long
standing problem of the explicite treatment of higher twist contributions
and their evolution \cite{twist}, the generalization to non-forward kinematics
\cite{SPD}, interpolating between the DGLAP
\cite{DGLAP} and the  ERBL \cite{ERBL} limiting cases.
The existence of this interpolation has been pointed out early in \cite{GR}.

The small $x$ behaviour of structure functions and related questions
have drawn the attention also to the relations between the Bjorken and the
Regge limits. The evolution in the latter asymptotics, as far as it
proceeds in the perturbative region, is represented by the BFKL equation
\cite{BFKL,Lev89,Levrep}.

In view of these topics a discussion of the Bjorken asymptotics invoking
concepts not aligned to the standard approach may be of interest.

We work out the effective action concept in analogy to the high energy
effective action proposed to investigate the Regge asymptotics
\cite{eff}. This allows to emphasize interesting
similarities.
We follow also the concepts developed in \cite{BuFKL}. The basic idea is
to formulate the factorization of short and large distance contributions
in terms of amplitudes and to determine the multi-parton $t$-channel
intermediate state. The partons represent here certain modes of the
underlying gluon and quark fields. The operator product expansion is not
the main tool in this formulation; however it leads to a natural and
simple basis of operators, the essential part of which are the
quasi-partonic operators. The scale dependence of the renormalized
operators appears due to the interactions of the exchanged partons,
being pair interactions in leading order represented by the so-called
non-forward evolution kernels.

The effective action is obtained from the QCD action by separating modes
and integrating over some modes. This action allows to obtain the parton
interaction kernels by a simple calculation. We represent the
interaction by hamiltonians involving (in leading order) two annihilation
and two creation operators of partons. The latter depend on the parton
type (gluon, quark flavour and chirality), parton helicity and on the
position on the light ray. The hamiltonian form is convenient for
investigating the symmetries of the exchanged multi-parton system.
In particular this formulation is intended to provide a framework to
analyze the integrability of the interaction.

The light ray position as the essential variable of operators in the
Bjorken limit appeared explicitely first in the approaches by Geyer,
Robaschik et al. \cite{GR} and by Balitsky and Braun \cite{BB}.

The parton interaction operators can be obtained alternatively as
external field effective vertices, where two type of external fields
are introduces describing the asymptotic interaction with the
currents of high virtuality and with the hadrons. This calculation can be
done analyzing the space-time field configurations only without 
transforming to momentum space.

From the very beginning conformal symmetry was underlying the ideas
about Bjorken scaling and its violation. It turned into a tool for
finding multiplicatively renormalized operators \cite{ERBL,Makeenko} and
for
relating the forward to the non-forward evolutions \cite{BuFKL,Muller}.
Combining conformal and supersymmetry leads to interesting relations
between different evolution kernels \cite{BuFKL}.
The non-forward evolution kernels have been reanalysed recently in
\cite{BGR}. Their reconstruction on the two-loop level from the results for
the forward case has been  considered in \cite{BFM}.

Integrability of the effective interaction in high energy QCD amplitudes
has been discovered by Lipatov \cite{LevPadua} in the Regge limit in the
case of multiple exchange of reggeized gluons and a similar
symmetry property  was expected in the Bjorken limit. The first examples
of higher twist evolutions tractable by integrability have been
studied by Braun, Korchemsky et al. \cite{BDKM}. Some
implications have also been considered in \cite{Belitsky}.

Discussing the Bjorken limit it may be convenient to have the amplitude
of a definite process in mind. The non-forward deeply virtual Compton
scattering off the proton , $\gamma^*(q_1) + p(p_1) \rightarrow
\gamma^*(q_2) + p(p_2) $, is an appropriate example (Fig. 1a):
\beqar
q_{1/2} = q^{\prime} - x_{1/2} p^{\prime}, \ \
q^{\prime 2} = 0, p^{\prime 2} = 0, p^{\prime } \approx p_1, s = 2
p^{\prime }  q^{\prime } , \\ \nonumber
-q_{1/2}^2 = Q_{1/2}^2 = x_{1/2} \ \ s.
\eeqar
The Bjorken limit corresponds to $s \rightarrow \infty $ with the
Bjorken variables $x_1$ and/or $x_2$ being of order $1$.

This process is generic in the sense that already  the two-parton
exchange contribution has the kinematics of non-vanishing longitudinal
momentum transfer which is encountered anyway in the subchannels of
multi-parton  exchange (higher twist) contributions. It is generic also
because one limiting case, $ x_1 = x_2$,  is directly related to ordinary
deep inelastic scattering and another limiting case, $x_1 = 0 , x_2 =
1$,  can be related by modifying the hadronic initial and final states to the
factorized matrix elements (light-cone wave functions, distribution
amplitudes) appearing in hard exclusive production.

 \section{Effective QCD action}

\subsection{Light-cone action}
\setcounter{equation}{0}

The Bjorken asymptotics of the amplitude is a sum over factorized terms,
where one factor (A) is determined by the large momentum scale $Q$ and
the other (B) by the hadronic momentum scale $m$, Fig. 1b.
 The $t$-channel intermediate states of partons should be understood and
defined in such a way that the corresponding exchange contribution is the
one of a (quasi) real multi-particle state to the unitarity relation
continued from the physical region of the $t$-channel. In this case the
two factors are (continued from mass shell) multi-particle amplitudes
having the valuable properties of gauge invariance and analyticity.


\begin{figure}[htb]
\begin{center}
\epsfig{file=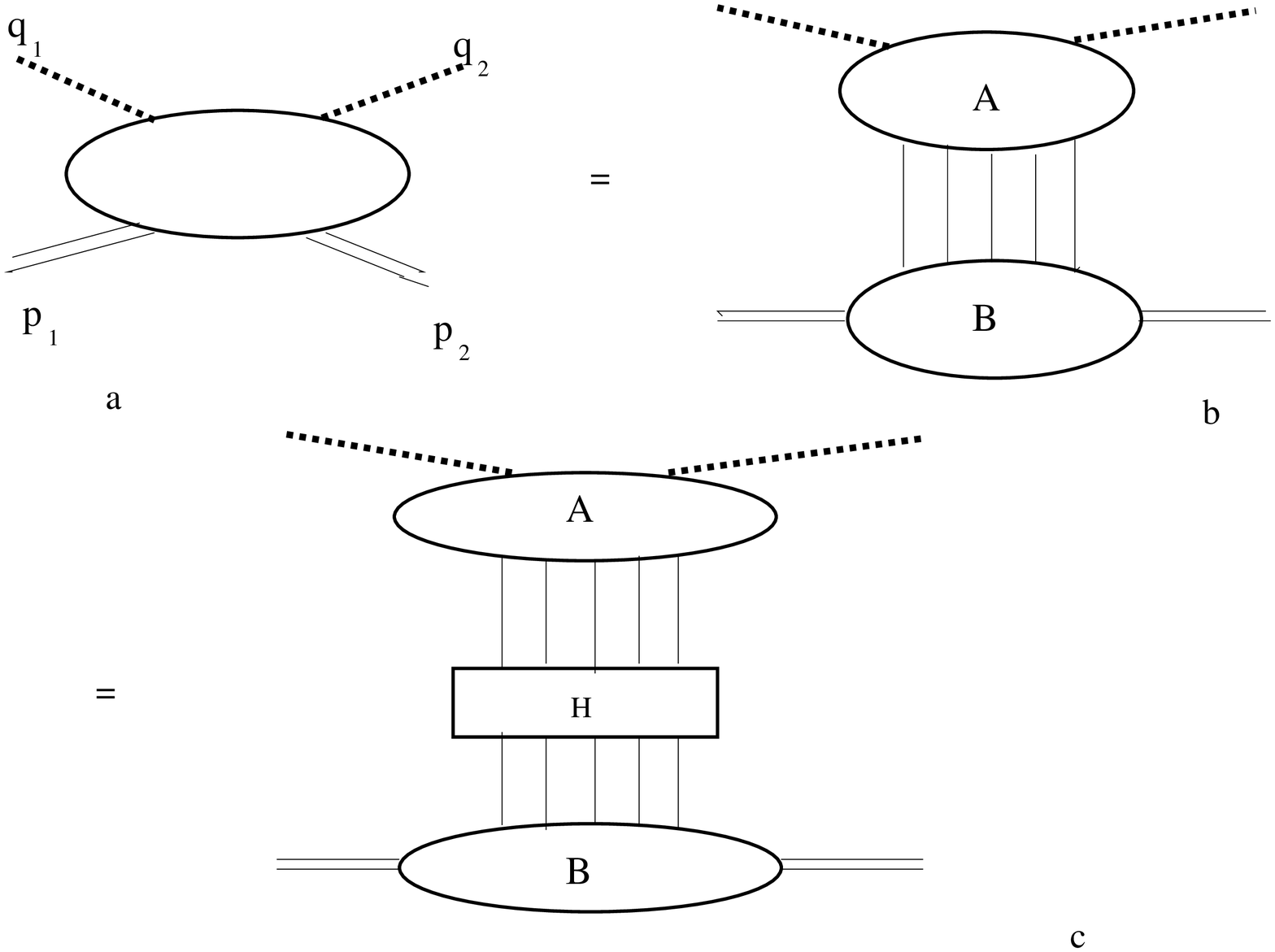,width=16cm}
\end{center}
\vspace*{0.5cm}
\caption{\small
Large-scale factorization of the amplitude. A summation over t-channel
intermediate states is implied.
}
\end{figure}

 The transverse momenta $\kappa_i$ of the exchanged partons are all of
the same order $\kappa, \ ( q \gg \kappa \gg m )$. The parton intermediate
states are to be specified further in such a way that the integration
over $\kappa_i$ results in $\ln {Q \over m}$, attributing the
non-logarithmic remainder to other (non-leading) exchanges.

The approach to the asymptotics can be visualized as a process of
iterating this factorization. The next step is shown in Fig. 1c. The
Green function $H$ (not necesserily connected) represents the parton
interaction. The effective action is to describe just this parton
interaction.

In the lowest order of perturbative expansion there are only pair
interactions. The partons can be identified as modes of the underlying
gluon $A$ and quark $ f, \tilde f$ fields in the gauge $q^{\prime \mu}
A_{\mu} = 0$ after integrating over the redundant field components
$p^{\prime \mu} A_{\mu}$ and $p^{\prime \mu} \gamma_{\mu} \ \psi $.

We represent 4-vectors $x^{\mu} $ by their light cone components
$x_{\pm} $ and a complex number involving the transverse
components $ x_{\perp} = x_1 + i x_2$.
 In the case of the gradient vector
we change the notation in such a way to have $\partial_+ x_- =
\partial_- x_+ = 1, \partial_{\perp} x = \partial_{\perp}^* x^* = 1$.
In particular the complex valued field $A$ represents the transverse
components of the gauge field. We choose the frame where the light-like
vector $q^{\prime}$ has the only non-vanishish component
$q^{\prime}_- = \sqrt {s/2}$ and $p^{\prime }$ the only non-vanishing component
$p_+^{\prime } = \sqrt {s/2}$.

We decompose the Dirac fields into light cone components,
\beqar
\psi = \psi_- + \psi_+, \ \ \gamma_- \psi_+ = \gamma_+ \psi_- = 0 \cr
\psi_+ = f u_{+-} + \tilde f u_{++}, \ \ \ \gamma^{\mu } = {2 \over s}
(\gamma_- q^{\prime \mu } + \gamma_+ p^{\prime \mu } ) + \gamma^{\mu }_{\perp }.
\eeqar

$u_{a,b}, \ \ a,b = \pm, $ is a basis of Majorana spinors,
\beq
\gamma_+ u_{-,b} = \gamma_- u_{+,b} = 0,  \ \
\gamma u_{a, -} = \gamma^* u_{a, +} = 0.
\eeq
In terms of component fields $A, f, \tilde f$ the QCD action can be written
as
\beqar
S \ = \ \int d^4 x  \{
 -2 A^{a *} (x) (\partial _+ \partial_- - \partial_{\perp} \partial_{\perp}^* )
A^{a} (x) +  \ \ \ \ \ \ \ \ \ \ \ \ \ \ \ \ \ \ \ \ \ \ \ \ \ \ \ \cr
 i f^{* \alpha } (x) \partial_+^{-1} (\partial_+ \partial_- - \partial_{\perp}
\partial_{\perp}^* ) f^{\alpha} (x) + ... \ \ \ \ \ \ \ \ \ \ \
\ \ \ \ \ \ \ \ \ \ \ \ \cr
 { g \over 2}( \partial_1 \hat V^*_{123} [ \partial_1 A^a (x_1) (A^*(x_2) T^a
A^*(x_3)) +  
  i f^{\alpha} (x_1) (f^*(x_2) t^{ \alpha} A^*(x_3) )
 + ...]_{x_i = x}  + {\rm c.c.} ) \cr
 + {g^2 \over 4 } \hat V_{1 1^{\prime}, 2 2^{\prime} }
[ 2 (A^*(x_1) T^c \partial A(x_{1^{\prime}})
(\partial A^*(x_2) T^c A(x_{2^{\prime}}) ) + \ \ \ \ \ \ \ \ \ \ \ \  \cr
 + \frac{1}{2} (f^*(x_1) t^c f(x_{1^{\prime }})) (f^*(x_2) t^c f(x_{2^{\prime
}}))  +    i  (f^*(x_1) t^c f(x_{1^{\prime }})) (A^* (x_2) T^c
\stackrel{\leftrightarrow} \partial A(x_{2^{\prime }}) )   \cr
  - i (f^*(x_1) t^{\alpha}  A(x_{1^{\prime }})) \partial  (A^*(x_2) t^{* \alpha}
f(x_{2^{\prime }}))
+ ... ]_{x_i = x_i^{\prime } = x } \     \}.
\label{action}
\eeqar
The periods stand for terms obtained from the written ones by replacing
pairs of fermion fields $f^*, f$ of one chirality by the ones of the other
chirality $\tilde f^*, \tilde f$ and by the pairs of other flavour fermion
fields.  The elimination of redundant field components has lead to non-local
vertices,
\beqar
\hat V^*_{123} = {i \over 3 \partial_1 \partial_2 \partial_3} [
\partial_{\perp 1}^* (\partial_2 - \partial_3) + {\rm cycl.} ], \ \
\hat V_{ 1 1^{\prime }, 2 2^{\prime }} = (\partial_1 +
\partial_{1^{\prime }})^{-2}.
\eeqar
Here and in the following we omit the space index $+$ on derivatives, i.e.
derivative operators not carrying subscripts $-, \perp $ are to be read as
$\partial_+$.  Integer number subscripts refer to the space point
on which the derivative  acts. The definition of the inverse $\partial^{-1}$ is to be
specified by the Mandelstam-Leibbrandt prescription.

The gauge group structure of the action has been written by using 
brackets combining two fields into the colour states of the adjoint ($a$)
and  of the two fundametal ($\alpha$ and $*\alpha$) representations:
\beqar (A_1^* T^a A_2 ) = -i f^{a b c } A_1^{* b} \ A_2^c, \ 
     (f_1^* t^a f_2) = t^a_{\alpha \beta} f^{* \alpha}_1 f_2^{\beta}, \cr
      (f^* t^{\alpha} A ) = t^c_{\beta \alpha} f^{* \beta} A^c, \ 
     (A^* t^{* \alpha } f) = t^b_{\alpha \beta} A^{* b} f^{\beta }.
\label{brackets} 
\eeqar

\subsection{Separation of momentum modes}

The partons in the intermediate state between the high   (low)  momentum scale
amplitude factor A (B)  and  H  are the large (+)  (small (-) ) transverse
momentum modes of the fields $A, f, \tilde f $. We have the mode
decomposition
\beq
A = A^{(+)} + A^{(0)} + A^{(-)}
\label{separation}
\eeq
(and analogously for the other fields). The medium modes  $A^{(0)} $ are
not important  in  the leading contribution discussed below. Medium modes
have to be separated in the leading contribution to the double logarithmic
approximation, compare Sect. 4.

The bare effective action of the leading parton interaction in the Bjorken
asymptotics is obtained from (\ref{action}) by substituting the mode separation  and
simplifying the
vertices by the approximation,
\beq
\partial^{\perp} A^{(-)} \ = \ 0, \ \ \partial^{\perp} f^{(-)} = 0, \ \
\partial^{\perp } \tilde f^{(-)} = 0.
\label{reduction}
\eeq
In particular the terms involving two fields in $(+)$ modes are the ones
important for deriving the bare $2 \rightarrow 2$ parton interaction
operators, as will be explained in an example in Sect. 3.

This effective action is the analogon of the high-energy effective action
considered in \cite{eff} as a tool for investigating the Regge limit in QCD.
Unlike the Regge case here the effective vertices are just the original QCD
vertices reduced by the conditions (\ref{reduction}).
 There is no analogon to the induced contribution to the triple vertices.

Going beyond the tree approximation one should notice that in this action
the modes of virtualness essentially larger than the scale $\kappa$ of the
(+) modes have been integrated out. In particular the coupling is the one at
scale $\kappa $, $\alpha_S(\kappa^2) $, and the the propagators of the (+)
modes are the bare ones at this scale. With the running coupling one defines the evolution
variable $\xi $,
\beq
\xi (\kappa^2, m^2) = \int_{m^2}^{\kappa^2} {d \kappa^{\prime 2} \over
\kappa^{\prime 2} }\ { \alpha_S( \kappa^2) \over 2 \pi }.
\label{xi}
\eeq

The propagators of the (-) modes, however, are normalized to be the bare
ones at the scale $\kappa^{\prime}, \kappa^{\prime } \ll \kappa $, appearing
in the next iteration. Therefore $Z$ factors of renormalization
\beq
Z_p = \exp (-  w_p \xi(\kappa^2, \kappa^{\prime 2}) ),
\eeq
where p labels the parton type, $p = A$ for gluon and $p = f $ for quark,
are to be included in the Green function $ H $ of parton interactions.
\beqar
w_p =  C_p \left ( \int_0^1 {d \beta \over 1- \beta } \ + \ w_p^{(0)} \right ),
\ \ \ \ C_A = N, \ \ C_f ={N^2 -1 \over 2N}, \cr
w_A^{(0)} =  -\frac{1}{4} ({11 \over 3} - {2 N_f \over 3 N} ),
\ \ \ \ w_f^{(0)} = - {3 \over 4}.
\label{wp}
\eeqar
The $Z$ factors refer to a finite change of renormalization scale. The parton
anomalous dimensions $w_p$ are infrared divergent (appearing as an artifact
of the gauge chosen). The divergence is written explicitely above without
specifying a regularization.

Comparing to the Regge case one understands that the $Z$ factors in the parton
propagators are the analogon  of the reggeization  factors $ ({s \over
m^2})^{\omega_p (\kappa ) }$ in the reggeon propagators of gluons or quarks in
the Regge case. The anomalous dimension $w_p$ is the analogon of the
trajectory function $\omega_p(\kappa)$. Also the point about the infrared
divergencies and their cancellation is analogous: In the same way  as in the
Regge case the infrared divergencies in the bare interactions are compensated
by the ones in $w_p$. The infrared cancellation in the multi-parton exchange
channel of overall colour singlet holds by the same arguments as in the
multi reggeon exchange \cite{BKP}:


\begin{figure}[htb]
\begin{center}
\epsfig{file=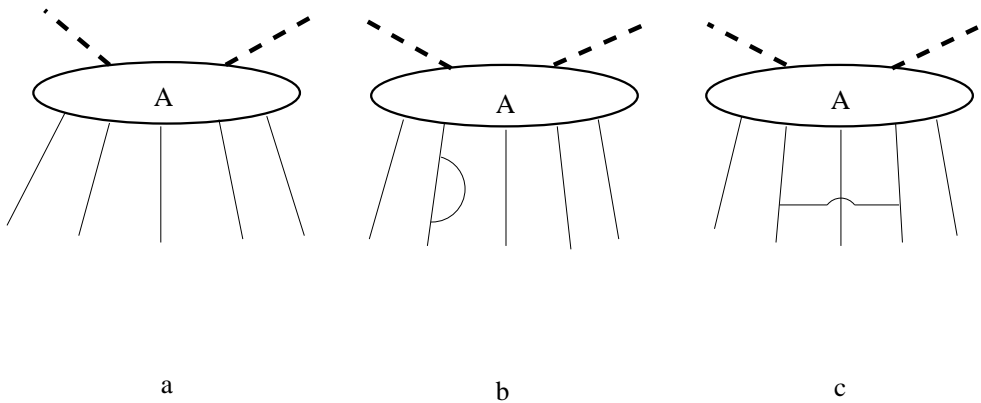,width=13cm}
\end{center}
\vspace*{0.5cm}
\caption{\small
Compensation of infrared singularities.
}
\end{figure}

The infrared singular contribution arising from vanishing gluon
(longitudinal in the Bjorken case and transverse in the Regge case) momentum
from the loop in Fig. 2c is twice the one in Fig. 2b, besides of the gauge
group factors. The relation of these group factors follows from the
observation that the exchanged partons or reggeons are in the overall colour
singlet state, i.e. a gauge group transformation is acting on the group
indices $a_1, ..., a_n$ (at this point the notation does not distingiuish
between the adjoint ($a$) and fundamental ($\alpha $) representations)
of  each of the factorized amplitudes as

\beq
\left (\exp[ i \sum \varepsilon^a T^a_i ] \right
 )^{a_1,...,a_n}_{a_1^{\prime}, ...,a_n^{\prime }}
A^{a_1^{\prime }, ...a_n^{\prime}} \ = \
A^{a_1, ..., a_n}
\eeq
The generators
$T_i^a, a=1, ..., N^2-1,$ act in the representation space of the
 parton
or reggeon  $p_i$. Expanding up to the second order in the group
parameters
$\varepsilon^a$ we find that in acting on the considered amplitudes one can
identify
\beq
\sum_i T_i^a \cdot T_i^a \ = \  -2 \sum_{i<j} T_i^a \otimes T_j^a,
\label{gaugerearrange}
\eeq
where $T_i^a\cdot T_i^a = I_i \ C_{p_i} $, with $I_i$ being the identity
matrix and $C_{p_i} $ the eigenvalue of the Casimir operator in the
representation corresponding to the parton $p_i$.

This allows to rewrite the singular contributions of Fig. 2b as
decomposed in terms of the
same gauge group factors as appearing in Fig. 2c and to perform the
cancellation of the infrared singularities explicitely.

The leading parton interaction is calculated by writing the   Green
function
$H  $ according to the effective action and extracting the logarithmic
contribution from the loop integrals of the (+) mode intermediate
state over
the transverse momenta $\kappa $ and the longitudinal (parallel  to
$q^{\prime }$
 momenta $\alpha$. These integrals are done in a standard way leaving
one-dimensional (in the longitudinal momentum $\beta$ parallel 
to $p^{\prime}$ )
loop integrals. In this way the parton interaction appears reduced
to one dimension, in the light ray
in space-time parallel to $q^{\prime }$ )
or in the light cone momentum (fraction) parallel to $p^{\prime }$.
Again this is analogous to the Regge case with an important difference: In
the Regge case the leading contribution is obtained by extracting the
logarithm from the longitudinal momentum integrations ($\alpha$ and $\beta
$) and therefore the reggeon interaction is reduced to the two transverse
dimensions.

It is well known that in the Regge case the most convenient way to deal
with the trivial longitudinal dimensions is to change to the continued
$t$-channel partial waves, i.e. to Mellin transform the amplitude factors
with respect to energies. In the same way it
turns out to be convenient to treat the trivial transverse momentum
dependence of the Bjorken limit amplitude factors by a Laplace transform
in $\xi$ ($\xi (Q^2, \kappa^2$)  in A, $\xi (\kappa^2, \kappa^{\prime
2}) $ in  H , $\xi (\kappa^{\prime^2}, m^2)$ in B). We perform this
transformation after changing the momentum variable $\alpha = {2 (k p) \over
s}$ to  $\bar \alpha = {\alpha s \over |\kappa |^2} $, because the
integrals in
$\bar \alpha $ are standard ones depending on momentum components $\beta 
= { 2 (kq^{\prime}) \over s}$ only.

The multi-loop convolution of the factorized parts A, H, B of the
amplitude reduces to a convolution in the longitudinal momentum fractions
$\beta $ only. The Laplace transforms of A,  H, B  enter at the same value
of the variable $\nu$ Laplace conjugated to $\xi$. The dependence on $\nu$ of
the leading contribution to the parton interaction  $H$  is just through the
factor ${1 \over \nu}$,
\beq
H (\alpha_i, \beta_i, \kappa_i; \alpha_j^{\prime}, \beta_j^{\prime },
\kappa_j^{\prime }) \rightarrow
{1 \over \nu } H^{(0)} (\beta_i; \beta_j^{\prime })
\eeq

The $Z$ factor contribution may be included at this point by the replacement
\beq
{1 \over \nu} \rightarrow  { 1 \over \nu + \sum_i w_{p_i} }
\eeq
The sum is over the exchanged partons of type $p_i$. This is the "energy
denominator" ($- \nu$ being the analogon of $E$ ) attributed to the
propagation of a multi-parton intermediate state between two subsequent
interactions.

The iterated factorization described above can be formulated as an
evolution equation in $\xi$ for the lower scale amplitude factor $B$.
\beqar
M_B(\beta_i) = M^{(0)}_B(\beta_i) + \ \ \ \ \ \ \ \ \ \ \ \ \ \ \ \cr
{1\over \nu + \sum w_p} \int d\beta^{\prime } \delta (\sum \beta_i - \sum
\beta^{\prime}_j ) \ \
H^{(0)} (\beta_i; \beta_j^{\prime }) \ M_B(\beta_j^{\prime })
\label{evoleq}
\eeqar
We have not written the gauge group and parton type indices carried by $M_B$
and $H$ and summed over in their contraction. The redefined kernels
\beq
H (\beta_i; \beta_j) = H^{(0)} (\beta_i; \beta_j^{\prime }) +
\prod_i^{n-1} \delta (\beta_i - \beta_i^{\prime}) \sum_i^{n} w_{p_i}
\eeq
represent infrared finite operators.
The rearrangement of the gauge group factors explained above
(\ref{gaugerearrange}) is implicite here.
The solutions of the evolution equation (\ref{evoleq}) is obtained in terms of a
complete set of eigenfunctions of these operators and their eigenvalues.
Usually one associates with the eigenfunctions composite operators,
being renormalized multiplicatively, and calls the corresponding eigenvalues
anomalous dimensions of these operators.

\subsection{Space-time picture}

The effective action describing the parton interactions in block H of
the virtual Compton amplitude can be alternatively treated from the
space-time point of view. Indeed, H describes the interaction between
sources located in the vicinity of the light ray $ x_{\perp} = 0, x_- =
0, x_+ = z \in {\bf R}^1 $ and other sources the distribution of which
is almost constant in the variables $x_{\perp}^{\prime}, x_+^{\prime}$
depending essentially only on the coordinate along the light-ray
$x_-^{\prime } = z^{\prime} $. The small width of the first distribution
near
the light cone is characterized by the short  distance scale $\Delta
\sim Q^{-1} $ and
the  variation of the second distribution in directions away from the
light ray is determined by the large distance scale $\delta \sim
m^{-1}$.

Consider now the QCD functional integral with such sources or the
related  vertex functional with corresponding external gluon and quark
fields.
Instead of separating momentum modes (\ref{separation})  
from this viewpoint we divide
now the fields of quarks and gluons into two types of external fields
$A^{(\pm) }$ and a quantum fluctuation $A_q$,
\beq A \rightarrow A^{(+)} + A_q + A^{(-)}.
\eeq
$A^{(+)} $ has the support near the light cone and has to be substituted
by the following expression in terms of (regularized) delta functions,
\beq A^{(+)} (x) = A_1 (z) \delta (x_+) \delta^{(2)} (x_{\perp}) +
{\cal O} (\Delta  ).
\label{fieldasymp+}
\eeq
The other external field $A^{(-)}$ has to be substituted as
\beq A^{(-)} (x^{\prime}) \  = \ A_1^{\prime } (z^{\prime}) \ {\rm const}
+
 {\cal O}( m ).
\label{fieldasymp-}
\eeq
The vertex functional or the external field effective action is now
obtained by doing   the integration over the quantum fluctuations $A_q$.
Consider in particular the resulting vertex involving two $A^{(+)}$ and
two $ A^{(-)}$ type fields on the tree level. It has the form
\beqar
\int d^4x_1 d^4 x_2  d^4x_{1^{\prime }} d^4 x_{^{\prime }}
A^{(+)}(x_1)  A^{(+)} (x_2)      G(x_{1 1^{\prime }} )
G(x_{2 2^{\prime}} )\cr
[\tilde V_{1 1^{\prime }} G(x_{1^{\prime} 2^{\prime }}) \tilde V_{2
2^{\prime }} + \delta^{(4) } (x_{1^{\prime} 2^{\prime }} )
V_{ 1 1^{\prime }, 2 2^{\prime }} ]  A^{(-)} (x_{1^{\prime} }) A^{(-)}
(x_{2^{\prime }})
\label{vertexansatz}
\eeqar

$\tilde V_{1 1^{\prime }}, \tilde V_{2 2^{\prime }} $ are simply related
to the triple vertex in (\ref{action}) depending on the 
case considered and  $G(x)$
stands for the quark or gluon propagator.
 We
substitute the particular asymptotic form of the external fields 
(\ref{fieldasymp+}, \ref{fieldasymp-}) and obtain
\beqar \int dz_1 dz_2 dz_{1^{\prime }} dz_{2^{\prime }}
A_1 A_2 A_{1^{\prime }} A_{2^{\prime}}
\ c({\delta \over \Delta }) \ \left
[\tilde V_{1 1^{\prime }} \tilde V_{2 2^{\prime }} \tilde J_{111} +
V_{ 1 1^{\prime }, 2 2^{\prime }} \tilde J_0 \right ]
\label{vertexz}
\eeqar
We abbreviate the residual dependence of the parton fields on the light
ray positions by indices $1,2$ for $A^{(+)}_1 (z_1)$, $ A^{(+)}_1 (z_2)
$ and by indices $1^{\prime }, 2^{\prime }$ for
$ A^{(-) \prime }_1 (z_{1^{\prime }}) $,
 $ A^{(-) \prime }_1 (z_{2^{\prime }}) $. The integration over the
transverse and $+$ components of $x_1, x_2$ is done due to the delta
functions and the integrals over the transverse and $+$ components of
$x_{1^{\prime}}, x_{2^{\prime }}$ are  represented by $\tilde J_{111},
\tilde J_0$,
\beqar
\tilde J_{111} = c(\delta / \Delta)^{-1} \int dx_{1^{\prime} -}
dx_{2^{\prime }-}
d^2 x_{1^{\prime } \perp} d^2 x_{2^{\prime } \perp } \partial^{\perp}_1
\partial^{ \perp * }_1  (x_{ 1 1^{\prime }}^2 )^{-1}
(x_{2 2^{\prime}}^2)^{-1} (x_{1^{\prime } 2^{\prime }}^2)^{-1}  \cr
\tilde J_{0} = c(\delta / \Delta)^{-1} \int dx_{1^{\prime} -}
dx_{2^{\prime }
-} d^2x_{1^{\prime } \perp} d^2 x_{2^{\prime } \perp } \partial^{\perp}_1
\partial^{ \perp * }_1  (x_{ 1 1^{\prime }}^2 )^{-1} (x_{2 2^{\prime
}}^2)^{-1} \delta^{(4)} (x_{1^{\prime } 2^{\prime }})
\eeqar
The integrals are regularized by taking into account the smearing of the
near light cone distribution by $\Delta $ and the large scale cutoff
$\delta$ for the other distribution. We shall do the integration in the
logarithmic approximation. For this we have substituted in 
(\ref{vertexz}) already
the effective triple vertices omitting terms which do not result in
logarithmic integrals.

In this approximation and after correcting the normalization  by a
factor $c(\delta / \Delta) \sim (\ln (\delta / \Delta))$
these integrals coincide with the Fourier transform of the standard $J$
integrals to be defined below (\ref{Jkernels}).

For going beyond the leading approximation one should specify accordingly
in more detail the asymptotic space-time dependence of the external
fields (\ref{fieldasymp+}, \ref{fieldasymp-}), include loop corrections
into the
vertices and perform the integrals beyond the logarithmic approximation.

In the above equations we did not
distinguish explicitely the cases of quarks and gluons
and we have suppressed the colour and helicity structures.
It turns out that these stuctures are the same for the $J_{111}$ and the
$J_0$ contributions. In this way the effective interaction vertices
factorize right from the first step of calculation into one acting on the
light ray positions and one acting on the helicity, chirality, colour
and flavour degrees of freedom.

The operators acting on the positions
arise as a sum of singular terms. The singularities  cancel
partially due to a relation between
the $J$ integrals involved. To cancel the remaining singularities and to
arrive at well defined operators the disconnected contributions  of
one-loop self-energy corrections have to be included.

Let us point out that in this framework it is readily understood that
in the approximation of leading twist and  leading logarithms
there are no vertices resulting in the change of the number
of partons in the exchange channel. 
 Consider for example the vertex of $A^{(+)} A^{(+)}
A^{(-)} $. There is a logarithmic contribution in the integration over
the coordinates besides of the light-ray ones; however in this term
$A^{(-)}$ enters with a transverse derivative and this vanishes in the
considered approximation (compare (\ref{fieldasymp-}) ). Going beyond this
approximation means introducing exchange partons of non-leading type.

We have outlined the calculation of  the vertex  of two-parton
interaction to
order $g^2$ in the logarithmic approximation. It is given by
(\ref{vertexz}) up to
the coefficient ${g^2 \over 8 \pi^2 } \ln({\delta^2 \over \Delta^2} ) $.
By renormalization group the result can be extended to all orders in the
leading logarithmic approximation. This amounts in replacing the kernel
in this vertex (\ref{vertexz} ) by the Green function of the Hamiltonian
operator
obtained from (\ref{vertexz}) by converting the external fields to
operators,
creation operators for the fields of type $A^{(+)}$ at points $1, 2$ and
annihilation
operators for the ones of type $A^{(-)}$  at $1^{\prime }, 2^{\prime }$.
The role of
(euclidean) time  in this Hamiltonian picture is played by the evolution
variable $\xi(\delta^2, \Delta^2 )  $ (\ref{xi}) , related to the
logarithmic coefficient
of the above vertex by $ d \xi = {g^2(\kappa^2) \over 8 \pi^2 }
{d\kappa^2 \over \kappa^2}  $, where the coupling ${g^2 \over 4 \pi} $ is
replaced by the (one-loop) running coupling $ \alpha_S (\kappa^2)$.

The spectrum of the Hamiltonian operator is actually the set of
anomalous dimensions appearing in the considered channel.

\section{Example of two-parton interaction}

\subsection{Energy-momentum representation}
\setcounter{equation}{0}

In order to illustrate the scheme described above we do the calculation for
the lowest order interaction of two gluons of parallel helicities.
The contribution of the triple vertex (\ref{action}) reduced to the effective vertex
by the condition (\ref{reduction}) to the graphs Fig. 3 are


\begin{figure}[htb]
\begin{center}
\epsfig{file=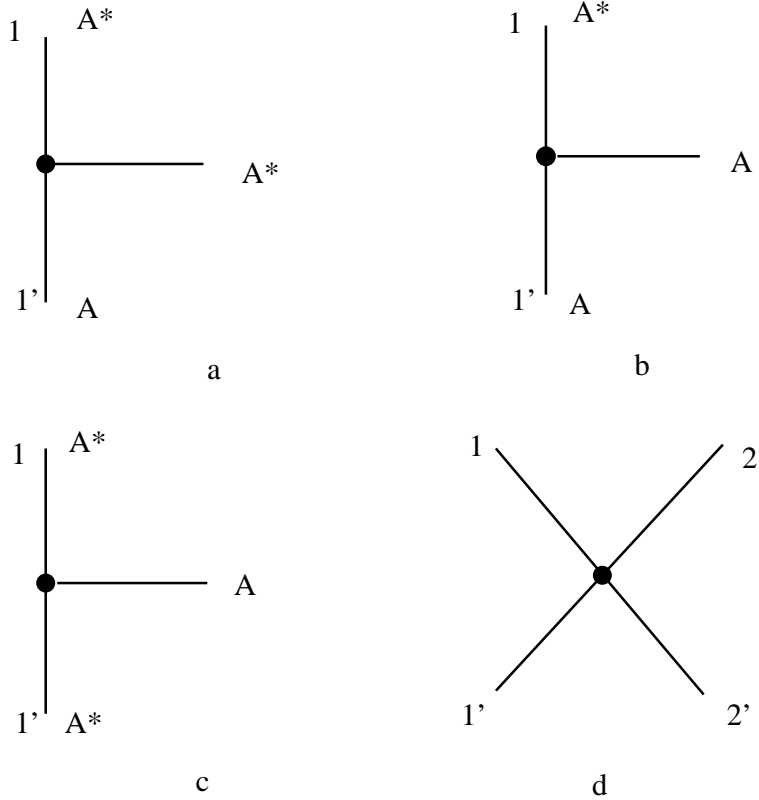,width=10cm}
\end{center}
\vspace*{0.5cm}
\caption{\small
Triple and quartic vertex graphs.
}
\end{figure}

\noindent
Fig. 3a : \hspace{6cm} Fig. 3b:
\beq
g T_{a_1 a_1^{\prime }}^c  {\beta_1^{\prime 2} \over \beta_1 -
\beta_1^{\prime} } {\kappa^*_1 \over \beta_1 }
\ \ \ \ \ \ \ \ \ \ \ \ \ \ \ \ \ \
g T_{a_1 a_1^{\prime }}^c  {\beta_1^2 \over \beta_1 -
\beta_1^{\prime} } {\kappa_1 \over \beta_1 }
\label{fig3ab}
\eeq
The quartic vertex in (\ref{action}) contributes to Fig. 3d as
\beq
- g^2 T^c_{a_1 a_1^{\prime }}  T^c_{a_2 a_2^{\prime }}
{ \beta_1 \beta_2^{\prime} - \beta_2 \beta_1^{\prime}
  \over (\beta_1 - \beta_1^{\prime })^2 }
\label{fig3c}
\eeq
We consider the contribution of these graphs to the two-parton interaction
Fig. 4a. There are three contributions as shown in Fig. 4b plus the crossing
contributions obtained by interchanging $1^{\prime }$ and $2^{\prime }$.
The first two terms differ by helicity of the gluon exchanged in
$s$-channel.

The expression for the graph Fig. 4a is ($k = \alpha q^{\prime } + \beta p +
\kappa $)
\beqar
   T^c_{a_1 a_1^{\prime }} T^c_{a_2 a_2^{\prime }}
{g^2 \over (2 \pi)^4 } \int {d^4k_1 \over (k_1^2+i \epsilon ) (k_2^2 + i
\epsilon) } \\ \nonumber
\left ( {\beta_1^{\prime 2} \beta_2^2 + \beta_1^2 \beta_2^{\prime 2} \over
(\beta_1 - \beta_1^{\prime } )^2 \beta_1 \beta_2 } \ 
{ |\kappa_1 |^2 \over (k_1 -k_1^{\prime })^2 + i \epsilon }
- {  (\beta_1 \beta_2^{\prime} + \beta_2 \beta_1^{\prime})
 \over (\beta_1 - \beta_1^{\prime } )^2 }
\right ) \ \
A^{a_1 a_2 } (q_1, q_2, k_1, k_2)
\eeqar
$A^{a_1 a_2}$ is the large-scale amplitude factor redefined to be
dimensionless by extracting powers of $Q$.
\beq
A^{a_1 a_2 } (q_1, q_2, k_1, k_2)
= \int_{-i\infty}^{+i\infty} {d \nu \over 2 \pi i}
A^{a_1 a_2 } (\nu, x_1, \beta_1, x_2, \beta_2) \ \
e^{\nu \xi(Q^2, \kappa^2)}
\eeq
We parametrize the momenta as $k_i = \alpha_i q^{\prime } + \beta_i p +
\kappa_i , \ \ \kappa_1 + \kappa_2 \approx 0$. The momenta of the (-)
modes are
to be substituted as $k_i^{\prime } = \beta_i^{\prime } p $. We change
$\alpha_1 \approx - \alpha_2 = \alpha$ by 
$\bar \alpha = { \alpha s \over |\kappa |^2 }$ and obtain
\beqar
\int_{m^2}^{Q^2} {g^2(\kappa^2) \ d \vert \kappa \vert^2 \over
16 \pi^2 \ \vert \kappa \vert^2 }
\int_0^1 d\beta_1 d\beta_2 \delta(\beta_1 + \beta_2 - \beta_1^{\prime }
-\beta_2^{\prime })
T^c_{a_1 a_1^{\prime } }
T^c_{a_2 a_2^{\prime } } \cr
\left \{ { \beta_1^{\prime 2} \beta_2^2 + \beta_1^2 \beta_2^{\prime 2}
\over (\beta_1 - \beta_1^{\prime })^2 \beta_1 \beta_2 }
J_{111}(\beta_1, -\beta_2, \beta_{1 1^{\prime} }) -
{ \beta_1 \beta_2^{\prime} + \beta_2 \beta_1^{\prime}
 \over (\beta_1 - \beta_1^{\prime })^2 } J_{11}(\beta_1, -\beta_2 )
\right \} \cr
A^{a_1 a_2 } (\nu, x_1, \beta_1, x_2, \beta_2) \ \
e^{\nu \xi(Q^2, \kappa^2)}.
\label{loopint}
\eeqar


\begin{figure}[htb]
\begin{center}
\epsfig{file=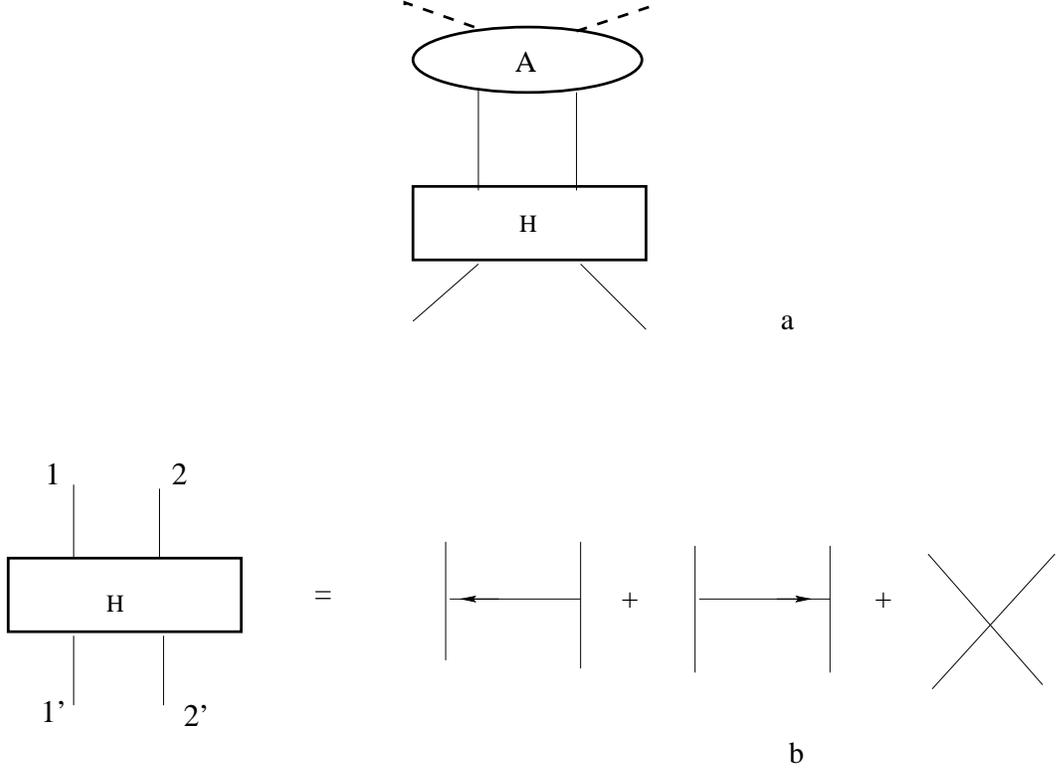,width=14cm}
\end{center}
\vspace*{0.5cm}
\caption{\small
Effective parton interaction.
}
\end{figure}

The $J$-functions represent standard $\bar \alpha$ integrals as
explained in detail in  Appendix A. The integral over $\kappa$ leads to
the factor ${1 \over \nu}$ and we identify the bare interaction kernel
$H^{(0)}$ as the expression in the braces. We use the first relation in
(\ref{relations})
between the $J$ functions to cancel the second order pole at $\beta_1 =
\beta_1^{\prime} $. We use the abbreviation for differences by double
indices like $\beta_{1 ^{\prime }} = \beta_1 - \beta_{1^{\prime }} $.
The remaining first order pole is the infrared
singularity being cancelled by the renormalization contribution $w_f$.
As the result we obtain the kernel
\beqar
 H^{a_1 a_2}_{a_1^{\prime }a_2^{\prime }}
(\beta_1, \beta_2; \beta_1^{\prime }, \beta_2^{\prime } ) =
- T^c_{a_1 a_1^{\prime } }
T^c_{a_2 a_2^{\prime } } \\ \nonumber
\left \{ \left [{\beta_1^2 \over \beta_1^{\prime} [\beta_{1 1^{\prime }}]_+ }
 J_{11}(\beta_1,\beta_{1 1^{\prime }} ) \right ] \ + \
 \left [{\beta_2^2 \over \beta_2^{\prime} [\beta_{2 2^{\prime }} ]_+ }
 J_{11}(\beta_2, \beta_{2 2^{\prime }} ) \right ] \ + \
2 w_g^{(0)} \delta (\beta_1 - \beta_1^{\prime }) \right \}
{\beta_1^{\prime} \beta_2^{\prime } \over \beta_1 \beta_2 }.
\label{momkernel}
\eeqar
Besides of illustrating the general scheme the example shows the great
simplification of the calculation relying on the effective action.
In going to higher loops the main point of effort will lie in the
improvement of the effective action, again similar to the Regge case
\cite{FL}.
Still there remains quite a number of two parton channels besides of the
one considered here. The calculation can be optimized further and
reduced to just three cases by supersymmetry.

\subsection{Space-time representation}

We outline how to treat the same example in the space-time
representation following now the scheme of section 2.3.
The vertex in the external field effective action $ A^{* (+)} (x_1)
A^{* (+)} (x_2) A^{(-)} (x_{1^{\prime }}) A^{(-)} (x_{2^{\prime }}) $
gets a contribution from the original quartic vertex  (\ref{action})
and
one from contracting two triple vertices by integrating over the quantum
fluctuation $A_q$.
The triple vertices in the Bjorken limit efective action
(\ref{action},\ref{reduction}) corresponding to Fig. 3a, b are

\beq - i g {\partial^2_{1^{\prime}} \partial_1^{* \perp} \over
(\partial_1 \partial_{1^{\prime}}) \partial_1 } A_q^{a *} \
(A_1^* T^{a} A_{1^{\prime}}) \vert_{x_1=x_{1^{\prime}} = x_q}; \ \ \
 i g {\partial^2_{1} \partial_1^{ \perp} \over
(\partial_1 \partial_{1^{\prime}}) \partial_1 } A_q^{a} \
(A_{1}^* T^{a} A_{1^{\prime }}) \vert_{x_1=x_{1^{\prime}} = x_q}
\label{triplex}
\eeq
Writing down the extenal field vertex as in (\ref{vertexansatz}),
substituting the
asymptotic form of the external fields and doing the interations besides
of the ones in the light-ray coordinates we obtain as the particular
case of (\ref{vertexz})
\beqar \int dz_1 dz_2 dz_{1^{\prime}} dz_{2^{\prime}}
\left [ {\partial_1^2 \partial_{2^{\prime}}^2 + \partial_{1^{\prime }}^2
\partial_2^2 \over (\partial_1 + \partial_{1^{\prime}} )^2 }
\tilde J_{111} +
 {\partial_1 \partial_{2^{\prime}} + \partial_{1^{\prime }}
\partial_2 \over (\partial_1 + \partial_{1^{\prime}} )^2 }
\partial_1 \partial_2 \tilde J_0  \right ]
\cr  (A_1^* T^{a} A_{1^{\prime}} )\
  (A_2^* T^{a} A_{2^{\prime}} )
\label{vertext1}
\eeqar
The first term in the brackets is  obtained by contracting the triple
vertices by substituting the points $1, 1^{\prime }$ in the second and
$2,2^{\prime } $ in the first vertex in (\ref{triplex}) and vice versa.
The
second terms emerges from the quartic vertex (\ref{action}) for the case
that
$A_1^*, A_2^*$ are of $A^{(+)} $ type and the remaining two field of the
$A^{(-)}$ type.

The second relation in (\ref{relations}) Fourier transformed to 
light ray variables implies
\beq {\partial_1^2 \partial_{2^{\prime}}^2 \over (\partial_1 +
\partial_{1^{\prime}})^2 } \tilde J_{111} \ + \
{ \partial_1^2 \partial_2  \partial_{2^{\prime}} \over (\partial_1 +
\partial_{1^{\prime}})^2 } \tilde J_{0} \ = \  \partial_{1^{\prime }}
\partial_{2^{\prime}} \ \tilde J_{1 1^{\prime }}^{(g)},
\eeq
where
\beq \partial_1 \partial_2 \tilde J_{1 1^{\prime }}^{(g)} =
- \int_0^1 d \alpha {(1 - \alpha)^2 \over \alpha } \ \delta(z_{1 1^{\prime
}} - \alpha z_{12}) \delta(z_{ 2 2^{\prime }})
\label{J11g}
\eeq
We use this relation and the one obtained by interchanging
$ 1, 1^{\prime } \leftrightarrow 2, 2^{\prime} $.
The remaining divergence signalled by ${1 \over \alpha} $ in the latter
integral  is cancelled after including the disconnected contribution of
order $g^2$ to the self energy in the two propagators connecting
$A_1^*$ with $A_{1^{\prime}}$ and
$A_2^*$ with $A_{2^{\prime}}$ separately,
\beq 2 \int dz_1 dz_2 dz_{1^{\prime }} dz_{2^{\prime }} \
w_g \ \delta (z_{1 1^{\prime }}) \delta (z_{2 2^{\prime }})
\  (A_1^* T^{a} A_{1^{\prime}} )\
  (A_2^* T^{a} A_{2^{\prime}} )
\label{disconnected}
\eeq
We have applied (\ref{gaugerearrange}) to obtain the above form seemingly
connected by colour interaction. $w_g$ is given by (\ref{wp}). Adding
(\ref{disconnected}) to (\ref{vertext1}) the
singular part of $w_g$ results in replacing in $\tilde J^{(g)}_{1 1^{\prime }}$
(\ref{J11g}) ${1 \over \alpha } $ by ${1 \over [ \alpha]_+ } $, adopting the
conventional "plus prescription". We obtain
\beqar
 \int dz_1 dz_2 dz_{1^{\prime}} dz_{2^{\prime}}
[ \tilde J^{(g)}_{11^{\prime }} + w_g^{(0) } \tilde \delta^{(2)} +
 (1, 1^{\prime } \leftrightarrow 2, 2^{\prime} ) ] \cr
 (A_1^* T^{a} \partial A_{1^{\prime}} )\
  (A_2^* T^{a} \partial A_{2^{\prime}} ), \ \ \ \ {\rm where } \ 
\partial_1 \partial_2 \tilde \delta^{(2)}  \ = \  \delta (z_{1 1^{\prime
}}) \  \delta (z_{2 2^{\prime }})
\label{vertext}
\eeqar
We have calculated in the space-time approach the vertex of parallel
helicity gluon interaction
to order $g^2$ in the logarithmic approximation. It is given by
(\ref{vertext}) up to
the coefficient ${g^2 \over 8 \pi^2 } \ln({\delta^2 \over \Delta^2} ) $.
As  noticed above by renormalization group the result can be extended
to all orders in the
leading logarithmic approximation.This leads to the Hamiltonian operator
obtained from (\ref{vertext}) by converting the external fields to
operators,
creation operators for the fields of type $A^{(+)}$ at points $1, 2$ and
annihilation
operators for the ones of type $A^{(-)}$  at the points
$1^{\prime }, 2^{\prime }$.

\section{Double-logartithmic gluon interaction}

\setcounter{equation}{0}

\subsection{Energy-momentum representation}

The scheme of factorizing the asymptotic amplitude into parts determined
by large and small scales and of extracting the logarithmic contribution
from the sum over intermediate states will be illustrated now in a
simpler situation. The combined asymptotics of large $Q^2$ and $s \gg
Q^2$ is of interest in deep-inelastic scattering at small $x \sim {Q^2
\over s}$. We avoid here the cases, where double logarithms appear in
the Regge asymptotics \cite{doublelog}. We consider the multiple gluon
exchange contributing to the amplitude (e.g. of virtual Compton
scattering at small $x_{1/2}$) behaving like $s^1$ on the tree level. In this
case the double logarithmic asymptotics can be approached equally well
from both the Regge or the Bjorken regions.

Now the loop integrations over the intermediate states in the factorized
amplitude Fig. 1,2 have a leading contribution logarithmic both in
longitudinal and in transverse momenta. Extracting this leading
contribution (and attributing the remainder to non-leading exchange
contributions) we obtain that the integrations in all dimensions are
standard ones. A double Laplace transform in $\ln {s \over Q^2} $
(conjugated variable $j = 1+ \omega $) and in $\xi(Q^2, m^2)$
(\ref{xi}) ( conjugated variable $\nu$) is appropriate now.

The dependence on the large scales is only via the product $r^2 = \xi
\cdot \ln{s \over Q^2}$ and in the transformed representation we have
the dependence only on the product $\rho^2 = \nu \omega$. Actually these
products are squares of two-dimensional Lorentz vectors, in particular
$\rho^2 = \nu \omega, \underline \rho = ({\omega + \nu \over 2}, {
\omega - \nu \over 2}) $. These vectors squared are the evolution variable
$r^2$ and its conjugate $\rho^2$. The evolution proceeds independent of
the "rapidity" $\sim \ln ({\xi  / \ln{s \over Q^2} }) $ inside the
"light cone".

The double logarithmic effective action is obtained from the underlying
QCD action by substituting in (\ref{action}) the mode separation
(\ref{separation}) and
reducing the vertices by the conditions
\beq
\partial_{\perp} A^{(-)} = 0, \ \ \ \partial (\partial^{-1} A^{(+)}) =
0.
\eeq
The latter means that a term involving $A^{(+)}$ is to be neglected
unless it is enhanced by its small longitudinal momentum in the
denominator. Here also medium modes $A^{(0)}$ have to be separated,
carrying transverse momenta $\kappa$ as large as the ones in the (+)
mode but longitudinal momentum fractions $\beta$ much larger than the
ones in the (+) modes and rather compatible with the ones in the (-)
mode.

In momentum representation the obtained effective action
implies that the quartic vertex gives no leading contribution and
the triple vertex is reduced to
\beq
g \partial_{\perp} \partial^{-1} A^{(+) a *} i ( A^{(0) *} T^{a}
\stackrel{\leftrightarrow} \partial A^{(-)} ) + {\rm c.c.}
\eeq
 Therefore in the double logarithmic limit the triple vertex graph
Fig. 5a vanishes and Fig. 5b and Fig. 5c result in the same expression
\beq
g f^{a_1 a_{1{\prime }} c}  \ {\beta_1^{\prime } \over \beta_1 } \ \
\kappa_1^* .
\eeq
 The only non-vanishing pair interactions arise for gluons with opposite
helicities, one interaction producing helicity flip and one without flip
in going from  $1^{\prime}, 2^{\prime}$ to 1, 2. Since the vertices
coincide both cases lead to the same expression (compare Fig. 4a)
\beqar
f^{a_1 a_1^{\prime } c} \ f^{c a_2 a_2^{\prime } } {g^2 \over (2\pi)^4}
\int {d^4 k_1 \over k_1^2 k_2^2 (k_1 - k_1^{\prime })^2 }
{\vert \kappa_1\vert^2 \beta_1^{\prime 2 } \over \beta_1^2 } \ \
A^{a_1 a_2} \left (\xi(Q^2, \kappa_1^2) \ln {\beta_1 s \over m^2} \right )
\eeqar
The calculation proceeds as in Sect. 3.1 for the first term in
(\ref{loopint})
involving $\vert \kappa_1 \vert^2$. In the double logarithmic limit
$J_{111}(\beta_1, -\beta_2, \beta_{1 1^{\prime }})$ reduces to $
{\beta_1 \over
 \beta_1^{\prime 2} }$ and we obtain in the transformed representation
$(\rho^2 = \omega \nu )$,
\beq
f^{a_1 a_1^{\prime } c} \ f^{c a_2 a_2^{\prime } }
{1 \over \rho^2} A^{a_1 a_2} (\rho^2) .
\eeq
The effective interaction depends on helicity and colour states only; it
does not act on longitudinal or transverse momenta.

For a closed formulation of the helicity dependence we introduce the
basis vectors $\left ( \matrix{1 \cr 0 \cr} \right )$ and $\left (
\matrix{0 \cr 1 \cr} \right )$ for the helicities carried by
$A^{(+) *}, A^{(-)}$ and $A^{(-) *}, A^{(+)}$, respectively, and express
the interaction operators in terms of Pauli matrices.
The spin-flip interaction
$\left ( \matrix{1 \cr 0 \cr} \right )\otimes \left ( \matrix{0 \cr 1
\cr} \right ) \rightarrow \left ( \matrix{0 \cr 1 \cr} \right ) \otimes
\left ( \matrix{1 \cr 0 \cr} \right ) $
 is described by the permutation
operator ${\cal P} = \frac{1}{2} (I\otimes I + \sigma_i^{(1)} \otimes
\sigma_i^{(2)})$ and the non-flip interaction
$\left ( \matrix{1 \cr 0 \cr} \right )\otimes \left ( \matrix{0 \cr 1
\cr} \right ) \rightarrow \left ( \matrix{1 \cr 0 \cr} \right ) \otimes
\left ( \matrix{0 \cr 1 \cr} \right ) $
by $- \sigma_3^{(1)} \otimes \sigma_3^{(2)}$. The latter is chosen such
that the sum of both operators describes also correctly the vanishing of
the interaction between parallel helicities
$\left ( \matrix{1 \cr 0 \cr} \right )\otimes \left ( \matrix{1 \cr 0
\cr} \right ) $.
The leading double logarithmic gluon pair Hamiltonian is
\beq
f^{a_1 a_1{\prime } c} \ f^{c a_2 a_2^{\prime } } \ \
 \frac{1}{2} \left (I\otimes I + \sigma_1^{(1)} \otimes \sigma_1^{(2)}
 +  \sigma_2^{(1)} \otimes \sigma_2^{(2)}
 -  \sigma_3^{(1)} \otimes \sigma_3^{(2)} \right )
\label{dloghamiltonian}
\eeq
We have reproduced a well known result to provide another example for
the effective action scheme. The formulation (\ref{dloghamiltonian})
 may be useful for
treating the double-logarithmic multi-gluon exchange by symmetry methods
of integrable systems.

\subsection{Space-time representation}

In the double logarithmic aymptotics the external fields should be
specified in the following way: The fields of type $A^{(+)} $
(interacting with the upper blob A in Fig. 1c) are concentrated in the
vicinity of the light ray and are almost constant in the light-ray
direction ($Q$ is small compared to $s$, $\delta_s \sim Q^{-1}$),
\beq \partial^{-1} A^{(+)} = (A^{(+)} + {\cal O}( Q) ) (\delta^{(2)}
(x_{\perp} ) \delta (x_+) + {\cal O} (\Delta) )
\label{dlogfield+}
\eeq
The fields of type $A^{(-)} $  (interacting with blob  B in Fig. 1c)
are almost constant in the directions pointing away off the light ray,
however the dependence on the light ray variable $z^{\prime} $ is a
narrow distribution of width $\Delta_s \sim s^{-1 } $ , approaching a
delta function,
\beq \partial A^{(-)} (x^{\prime }) \ = \  (A^{(-)} 
\delta(z^{\prime} - z_0)  + {\cal O}(\Delta_s )) \ 
({\rm const} + {\cal O} (m) )
\label{dlogfield-}
\eeq
$A^{(\pm)} $ on the right-hand sides of (\ref{dlogfield+},\ref{dlogfield-})
are constants;
they should not be identified with similar symbols in other sections.

Consider now the vertex $A^{(+)} (x_1) A^{(+)} (x_2) A^{(-)}
(x_{1^{\prime }} ) A^{(-)} (x_{2^{\prime }}) $. The integrations over
the coordinates pointing away off the light ray are done in logarithmic
approximation as above. Unlike the Bjorken asymptotics now also the
dependence on the light ray
coordinates is specified in the asymptotic external fields and therefore
also the integration over these variables can be done. We shall do these
integrations approximately extracting the second logarithm of the large
scale ratio.
We rely on the results of the Bjorken asymptotics for the gluon
interaction of parallel helicity (\ref{vertext1}) and of antiparallel
helicity
given in Appendix B (\ref{vertexa1}).  Now we use the information about the $z$
dependence specified in (\ref{dlogfield+},\ref{dlogfield-}) and notice that
the derivatives $\partial_1, \partial_2$ act
on (almost) constant fields. We conclude immediately that the
considered external field vertex
vanishes in the case of parallel helicities (\ref{vertext1} ) and in
the antiparallel helicity case (\ref{vertexa1}) it reduces to
\beqar \int d z_1 dz_2 d z_{1^{\prime }} dz_{2^{\prime }}
\delta (z_{1^{\prime}} - z_0)
\delta (z_{2^{\prime}} - z_0)  \partial_1 \partial_2 \tilde J_{111}  \cr
[ (A^{(+)*} T^{a} A^{(-)} ) (A^{(+)} T^{a} A^{(-) *} ) +
 (A^{(+)*} T^{a} A^{(-) *} ) (A^{(+)} T^{a} A^{(-) } ) ]
\label{dlogvertex}
\eeqar
We extract the logarithmic contribution in the integral over $\tilde
J_{111} $ for $z_{1^{\prime}}, z_{2^{\prime}} \rightarrow z_0 $
appearing as a divergence $\sim \alpha_3^{-1} $ in the $\alpha $
representation (\ref{relations}, \ref{Jkernels}).

Up to the coefficient $ \sim g^2( \ln(\delta / \Delta) ) \ln (\delta_s /
\Delta_s) $ the resulting vertex is given by the bracket in
(\ref{dlogvertex} ) involving
the double-logarithmic "fields", i.e. the constats $A^{(\pm)}$ carrying
merely the colour and helicity degrees of freedom.
In order to extend the result to all orders in the double-logarithmic
approximation we convert the obtained vertex into an Hamiltonian
operator by defining the "fields" $A^{(\pm)}$ to be creation and
annihilation operators. The resulting operator is of course equivalent
to the result obtained  in the momentum calculation
(\ref{dloghamiltonian}). From the latter
being a matrix representation  we recover  the operator representation
(\ref{dlogvertex}) by
multiplication by the helicity state vectors composed of the creation
and annihilation operators, $(A^{* (+)} , A^{(+)} )_1 \otimes ( A^{*
(+)},
A^{(+)} )_2 $ from the left and $ (A^{(-)}, A^{* (-)} ) _1^T \otimes
(A^{(-)}, A^{* (-)} )_2^T $ from the right.

 \section{Interaction operators in the Bjorken limit}
\setcounter{equation}{0}

The momentum kernels or the external field vertices calculated above
in the tree approximation are extended to all orders in the leading
logarithmic approximation. This extension is represented in terms of the
 evolution in "euclidean time" $\xi(Q^2, m^2)$, describing the
approach to the Bjorken asymptotics, which can be formulated in Hamiltonian
operator language. Let the parton fields $A^{(-)}, f^{(-)}, \tilde
f^{(-)}$ act as annihilation operators and $A^{(+)}, f^{(+)}, \tilde
f^{(+)}$ as creation operators with the contractions (longitudinal momentum
representation)
\beqar
<0| A^{a (-)} (\beta) A^{b (+) *}(\beta^{\prime }) |0> =
{1 \over 2} \delta (\beta - \beta^{\prime } ) \delta^{a b}, \cr
<0| A^{a (-)} (\beta) A^{b (+) }(\beta^{\prime }) |0> = 0, \cr
<0| f^{\alpha (-)} (\beta) f^{\beta (+) *}(\beta^{\prime }) |0> =
\beta  \delta (\beta - \beta^{\prime } ) \delta^{\alpha \beta}.
\eeqar
This corresponds to the following kinetic terms (light-ray position
representation)
\beq
\int dz \{ 2  (A^{a *(+)} A^{a (-)}) - i f^{\alpha (+)*} \partial^{-1}
f^{\alpha (-)}
+ {\rm c.c.} + ... \}
\label{hkin}
\eeq
The periods stand for the terms of other fermion chirality and flavours.
${\rm c.c.}$ does not exchange (+) and (-) and acts as conjugation
otherwise (normal ordering).
The parton interaction Hamiltonian operators  can be obtained from the
momentum kernels and also from the external field vertices,
\beqar
\int d^4\beta \delta(\beta_1 + \beta_2 - \beta_{1^{\prime }} -
\beta^{\prime } ) (-\beta_1 \beta_2)^{-1})
(A^{(+)*}(\beta_1) T^{a}A^{(-)}(\beta_{1^{\prime }})) \ \cr
(A^{(+)*}(\beta_2) T^{a} A^{(-)}(\beta_{2^{\prime }})) \ \
H (\beta_1, \beta_2; \beta_{1^{\prime}}, \beta_{2^{\prime }}) = \cr
\int d^4z
( A^{(+)*}(z_1) T^{a} A^{(-)}(z_{1^{\prime }})) \
( A^{(+)*}(z_2) T^{a} A^{(-)}(z_{2^{\prime }})) \
\tilde H (z_1, z_2; z_{1^{\prime}}, z_{2^{\prime }}).
\eeqar


There is a convenient representation  of the kernels
 in terms of Feynman parameter integrals. As shown in Appendix A,
the standard $\bar \alpha$ integrals $J(\beta,...)$ can be written in
Feynman parameter form allowing to do the Fourier transformation rather
easily.

We list the operators of the leading parton interaction in the QCD
Bjorken limit, the momentum kernels of which are well known,  in the
light-ray and Feynman parameter form. We restrict
ourselves to the ones involving gluons ($A$) and/or only one flavour and
one chirality of quarks ($f$). We use the abbreviations (compare
Appendix A)
\beqar
\partial_{1} \partial_{2}
\tilde J_{1 1^{\prime}}^{(p)} = \int_0^1 {d\alpha \over [\alpha]_+ }
\ \chi_1^{(p) }(\alpha) \  \delta (z_{1 1^{\prime}} - \alpha z_{12})
  \delta(z_{2 2^{\prime }}), \cr
\chi_1^{(0)} = \alpha (1 - \alpha),  \chi_1^{(f)} = - (1 - \alpha) ,
\chi_1^{(g)} = - (1-\alpha )^2, \cr
\partial_{1} \partial_{2}
\tilde J_{n_1 n_2 n_3} = {\Gamma(n_1+n_2+n_3-1) \over \Gamma(n_1)
\Gamma(n_2) \Gamma(n_3) } \int_0^1 d\alpha_1 d\alpha_2 d\alpha_3 \delta(\sum
\alpha_i)  \cr
\ \alpha_1^{n_1} \alpha_2^{n_2} \alpha_3^{n_3}
\  \delta (z_{1 1^{\prime}} - \alpha_1  z_{12})
 \delta(z_{2 2^{\prime } } +  \alpha_2 z_{12} ), \cr
\partial_{1} \partial_{2}
\tilde J_0^{(g)} = \int_0^1 d\alpha \ \chi_0^{(p)}(\alpha )  \
 \delta (z_{1 1^{\prime}} - \alpha  z_{12})
 \delta(z_{2 2^{\prime } } +  (1- \alpha) z_{12} ),
\cr
\chi_0^{(0)}  = 1, \ \chi_0^{(g)} = \alpha (1 - \alpha ), \ \ \ \
\partial_{1} \partial_{2}
\delta^{(2)} = \delta (z_{1 1^{\prime }}) \delta (z_{2 2^{\prime }}).
\label{Jkernels}
\eeqar
The case of parallel helicity gluon interaction has been done as the example
in section 3. In the cases of antiparallel helicity interactions some further
transformations of $J$ kernels (\ref{relations}) are applied for presenting
the result in a convenient form. In Apprendix B the case of gluons 
of antiparallel helicities is treated and a transformation applied in the
cases of  quark-gluon and annihilation type interactions is
given. 
 
The position arguments of the fields will be abbreviated as indices 1,
$1^{\prime }$, 2, $2^{\prime }$. We suppress the label $(\pm)$ distinguishing
creation and annihilation operators, since the fields at points $1, 2$ act
always as creation and the fields at the points $1^{\prime }, 2^{\prime }$
always as annihilation operators.
The integration over the positions,
summation over colour indices and operator normal ordering is implied.

\noindent
parallel helicity interactions
\beqar
 \{ 4 [\tilde J_{1 1^{\prime }}^{(g)} + w_g^{(0)} \delta^{(2)} ]
(A^*_1 T^{a} \partial A_{1^{\prime}})
-2 i [\tilde J_{1 1^{\prime }}^{(f)} + w_f^{(0)} \delta^{(2)} ]
(f_1^* t^{a} f_{1^{\prime}} ) \} \cr
[-2 (A_2^* T^{a} \partial A_{2^{\prime}} ) + i  (f_2^* t^{a}
f_{2^{\prime}} ) ] \cr
-4i  \tilde J_{ 1 1^{\prime }}^{(0)}
 (f_1^{*} t^{\alpha} A_{1^{\prime}})
\ ( A_2^* t^{* \alpha}  f_{2^{\prime }})   + {\rm c.c.}
\label{ttgf}
\eeqar
anti-parallel gluon interactions
\beqar
\{ 8 [ \tilde J_{1 1^{\prime }}^{(g)} + w_g^{(0)} \delta^{(2)} ]
\ + 4  \tilde J_{221} - 8 \tilde J_{112} \} \cr
 ( A_1^{*} T^{a} \partial A_{1^{\prime}})
( A_2 T^{a} \partial A^*_{2^{\prime }})   \cr
+ 4  \tilde J_{221}
 ( A_1^{*} T^{a} \partial A_{1^{\prime}}^*)
( A_2 T^{a} \partial A_{2^{\prime }})   + {\rm c.c.}
\label{hgg}
\eeqar
anti-parallel helicity quark interactions (one chirality, one flavour)
\beqar
\{-2  [\tilde J_{1 1^{\prime }}^{(f)} + w_f^{(0)} \delta^{(2)} ]
+ \tilde J_{111} \}
 (f_1^* t^{a} f_{1^{\prime}} )
 ( f_2 t^{a} f_{2^{\prime }}^*)   \cr
+ 2 \tilde J_0^{(g)}
(f_1^* t^{a}
 f_{2})
 (f_{1^{ \prime}} t^{a} f_{2^{\prime }}^*) + {\rm c.c.}
\label{hff}
\eeqar
anti parallel helicity quark - gluon interactions
\beqar
-4i [\tilde J_{1 1^{\prime }}^{(g)} + w_g^{(0)} \delta^{(2)} ]
 ( A_1^{*} T^{a} \partial A_{1^{\prime}})
 \ ( f_2 t^{a} f_{2^{\prime }}^*)   \cr
-4i  [\tilde J_{1 1^{\prime }}^{(f)} + w_f^{(0)} \delta^{(2)} ]
 ( f_1^* t^{a} f_{1^{\prime}})
( A_2 T^{a} \partial A^*_{2^{\prime }})  \cr
 - 4i  \tilde J_{211}
 [ ( f_1^* t^{  \alpha } \partial A_{1^{\prime}}^* ) \
  ( A_2 t^{ * \alpha} f_{2^{\prime }})
- (A_1^* T^{a}   \partial A_{1^{\prime }} )
(f_2 t^{a} f_{2^{\prime }}^* )  ] \cr
-8i  \tilde J_{111}
 (A_1^* T^{a}   \partial A_{1^{\prime }} )
(f_2 t^{a} f_{2^{\prime }}^* ) \ \
  + {\rm c.c.}
\label{hgf}
\eeqar
annihilation-type interactions
\beqar
 2i \tilde J_{1 1^{\prime }}^{(0)}
 [ (A_1^* t^{ *  \alpha}
  \partial f_{1^{\prime }})
 ( A_{2} t^{ \alpha} f_{2^{\prime }}^*) +
 ( f_1^* t^{ \alpha}
 \partial  A_{1^{\prime }})
 \ (f_{2} t^{* \alpha}  A_{2^{\prime}} ) ]  \cr
+ 2i   \tilde J_0^{(g)}
[ ( A_1^* T^{a}
\stackrel{\leftrightarrow} \partial   A_{2})
 (f_{1^{\prime}} t^{a} f_{2^{\prime }}^*)
+  ( f_1^* t^{a}
   f_{2})
 \ (A_{1^{\prime}} T^{a} \stackrel{\leftrightarrow}
\partial A_{2^{\prime }}^* )                   \cr
+ (\partial_{1^{\prime }} + \partial_{2^{\prime }})
[ (f_1^* t^{\alpha} A_{1^{\prime }}^*)  (f_2 t^{*\alpha} A_{2^{\prime }} )
\cr 
-  (f_1^* t^{\alpha} A_{1^{\prime }} )  (f_2 t^{*\alpha} A_{2^{\prime }}^* ) 
+ (A_1^* t^{\alpha} f_{1^{\prime }}^* ) (A_2 t^{*\alpha} f_{2^{\prime }} )
 ]] \cr
- 2i    \tilde J_{111}
 [  ( f_1^* t^{ \alpha}  \partial A_{1^{\prime }}^*)
 (f_{2} t^{ * \alpha}  A_{2^{\prime }} )
 +   ( A_1^* t^{ \alpha}
  f_{1^{\prime }}^*)
 ( A_{2} t^{ * \alpha} \partial  f_{2^{\prime }} ) ]          \cr
- 2i    \tilde J_{211}
 [  ( f_1^* t^{ \alpha}\partial   A_{1^{\prime }}^*)
 (f_{2} t^{ * \alpha}  A_{2^{\prime }} )
 +   ( A_1^* t^{ \alpha}\partial   f_{1^{\prime }}^*)
 (A_{2} t^{ * \alpha}  f_{2^{\prime }} )           \cr
-   ( f_1^* t^{ \alpha}\partial  A_{1^{\prime }})
 ( f_{2} t^{\alpha}  A_{2^{\prime }}^* ) ]
 + {\rm c.c.}
\label{hannih}
\eeqar

\section{Symmetries of the parton interactions}
\setcounter{equation}{0}

\subsection{Conformal symmetry}

It is well known that the leading order parton interactions are
conformally symmetric. In the present light-ray formulation this
symmetry is manifest. The interaction operators (\ref{ttgf} -  \ref{hannih})
 as well as the
kinetic operators (\ref{hkin}) are invariant under the transformations acting on
the light-ray positions as
\beq
z \rightarrow \tilde z = {az+b \over cz+d}
\label{confz}
\eeq
($a,b,c,d $ real) and on the fields as follows.
The fermionic parton fields $f^{(\pm)}$ appear as conformal primaries of
weight 1,
\beq f \rightarrow {1 \over ( cz+d)^2} f.
\label{conffields}
\eeq

In the case of gluons the operators $ \partial A^{(-)} $ and
$\partial^{-1} A^{(+)}$ appear as conformal primaries of weights
$\frac{3}{2}$ and $-\frac{1}{2}$.

Since translation and scale symmetries are obvious only the infinitesimal
proper-conformal transformations ($a=d=1, b=0, c $ infinitesimal)
are to be checked. 
Let us do the example of quarks of parallel helicity. The essential term
can be written  (in simplified notation disregarding colour) as
\beqar
\int d^4z \int_0^1 d\alpha {1-\alpha \over \alpha} \delta(z_{1
1^{\prime}} - \alpha z_{12} ) \delta (z_{2 2^{\prime }})
\partial^{-1} f^{(+)*}(z_1) \partial^{-1} f^{(+)*} (z_2)\cr
f^{(-)}(z_{1^{\prime}}) f^{(-)}(z_{2^{\prime}})
\label{ffop}
\eeqar
The transformation acts on the fields and results in

\beqar
\int d^4z \int_0^1 d\alpha {1-\alpha \over \alpha} \delta(z_{1
1^{\prime}} - \alpha z_{12} ) \delta (z_{2 2^{\prime }})
\partial^{-1} \left ({1 \over (1+cz_1)^2} f^{(+)*}(\tilde z_1) \right )
\cr
 \partial^{-1} \left ({1 \over (1+cz_2)^2} f^{(+)*} (\tilde z_2)\right )
 {1 \over
(1+c z_{1^{\prime }})^2 } f^{(-)}(\tilde z_{1^{\prime}})
{1 \over (1+cz_{2^{\prime }})^2} f^{(-)}(\tilde z_{2^{\prime}})
\label{fftrans}
\eeqar
We have to check that the latter two expressions coincide up to $ {\cal
O}(c^2)$. We have
 $\partial_{\tilde z} = (1+cz)^2 \partial_z, \ d\tilde z
= (1+cz)^{-2} dz $ and
\beqar
 \int_0^1 d\alpha {1-\alpha \over \alpha} \delta(z_{1
1^{\prime}} - \alpha z_{12} ) \delta (z_{2 2^{\prime }}) =
(1+c z_1)^{-2} (1+ c z_2)^{-2}   \cr
 \cdot \int_0^1 d\alpha {1-\alpha \over \alpha} \delta(\tilde z_{1
1^{\prime}} - \alpha \tilde z_{12} ) \delta (\tilde z_{2 2^{\prime }})
\eeqar
By substituting these relations we see that (\ref{fftrans}) transforms into
(\ref{ffop}) with the variables $z$ changed to $\tilde z$.

In this way we can check the conformal invariance of the remaining
interaction terms. In the form given above each term in (5.5-8) is
invariant separately. In the form (\ref{hannih}) for the annihilation-type
interaction this does not apply. This is related to the missing or extra
derivatives which can be traded, however, for powers of $z_{12} $ as
outlined in Appendix B. The annihilation type interaction terms 
(\ref{hannih}) can be written as
\beqar
-4i \tilde J_{111}^- \ (A_1^* t^{\alpha} f_{1^{\prime }}^* ) (A_2
t^{*\alpha } f_{2^{\prime }} )
\ + \ 2i \tilde \delta^{(2) -} \
 (A_1^* t^{* \alpha} f_{1^{\prime }} ) (A_2
t^{\alpha } f_{2^{\prime }}^* ) \cr
+ 12 i \tilde J_0^{(g)-} \ (A_1^* T^{a} A_2) (f_{1^{\prime }} t^{a}
f_{2^{\prime }}^* )  \cr
- \frac{2}{3} i \tilde J_{221}^+ \
[ (f_1^* t^{\alpha} \partial A_{1^{\prime }} ) \ (f_2 t^{*\alpha }
A_{2^{\prime }}^* )
-
 (f_1^* t^{\alpha} \partial A_{1^{\prime }}^* ) \ (f_2 t^{*\alpha }
A_{2^{\prime }} ) ]  \cr
-i \tilde J_{112}^+
[ (f_1^* t^{\alpha} \partial A_{1^{\prime }} ) \ (f_2 t^{*\alpha }
A_{2^{\prime }}^* )
\label{annihc}
\eeqar
Here the kernels with the additional subscript $\pm$ are related to the
corresponding ones used so far by
\beq \partial_1 \partial_2 \ \tilde J^{\pm} \ = \ z_{12}^{\pm} 
\partial_1 \partial_2 \tilde J.
\eeq 

Each term in this expression (\ref{annihc}) is conformally symmetric.

\subsection{Helicity symmetry}

The operators (\ref{ttgf} - \ref{hannih})
are invariant also under the $SU(1,1)$
transformation acting on helicities,  on  fields of type $A^{(+)}$,
the position argument of which we agreed to denote by $1$ or $2$,
\beq
\left ( \matrix{A_1^* \cr A_1 \cr} \right )
 \rightarrow \left ( \matrix{\tilde A_1^* \cr \tilde A_1 \cr} \right ) =
\hat V(a_1, a_2, a_3) \ \left ( \matrix{A^* \cr A \cr} \right ).
\eeq
where
\beq \hat V(a_1, a_2, a_3) \ = \
\exp [\frac{1}{2} (a_1 \sigma_1 + a_2 \sigma_2 -i a_3 \sigma_3]
\eeq
The fields of type $A^{(-)}$ transform analogously with $\hat V$
replaced by
$\hat V^{+ -1}$. In particular the light ray kinetic terms (\ref{hkin})
are left invariant.

The products of a field of type $A^{(+)}$ and one of type $A^{(-)} $
(say at positions 1 and $1^{\prime}$ )
decompose into the scalar and vector components, where the (2+1) vectors
of this group can be constructed in two ways, using
$(\sigma_r) = (\sigma_1, \sigma_2; \sigma_3)$  or using
$(\tilde \sigma_r) = (\sigma_1, \sigma_2; -  \sigma_3)$.
The two types of vectors $v$ and $\tilde v$  are related to each other
by a parity transformation, analogous to time reversal. In particular
the vectors constructed with $\frac{1}{2} (\sigma_r \pm \tilde
\sigma_r)$ are even or odd with repect to that parity. The subspaces of
even or odd vectors are two- or one-dimensional.

Thus the  four states in the tensor product $\left ( \matrix{A_1^* \cr
A_1
\cr} \right ) \otimes \left ( \matrix{A_{1^{\prime }}^* \cr A_{1^{\prime }}\cr} \right )$
decompose according to
\beqar
R = (A_1^* \ \  A_1) \hat R
\left ( \matrix{A_{1^{\prime}}^* \cr A_{1^{\prime }} \cr} \right ) ,
R = {\cal V}, {\cal A}, {\cal T}^{\pm} \eeqar
into the scalar representation,
\beq \hat R = \hat {\cal V} = \sigma_1,  \ \ R = {\cal V} = A_1^*
A_{1^{\prime} } + A_1 A_{1^{\prime }}^* ,
\eeq
the one-dimensional odd-vector representation,
\beq \hat R = \hat {\cal A} = \sigma_1 \sigma_3 , \  \ R = {\cal A} =
A_1^* A_{1^{\prime }} -  A_1 A_{1^{\prime}}^{*} ,
\eeq
and the two-dimensional even-vector representation,
\beq \hat R = \hat {\cal T}^{\pm}  = \sigma_1 \sigma_{\pm} , \  {\cal
T}^+ = A_1^* A_{1^{\prime}}^* ,  \ {\cal T}^- =   A_1 A_{1^{\prime }}.
\eeq
If the indices 1 and $1^{\prime}$ refer to out (+) and in (-) going
partons, then the product representations describe transitions, where
one has

${\cal V}_s$ : no helicity flip, helicity independent
($\delta_{\lambda \lambda^{\prime }}$ ),

${\cal A}_s$: no helicity flip, helicity dependent,
($\lambda \delta_{\lambda \lambda^{\prime }}$),

${\cal T}^{\pm}_s$: helicity flip from $\pm$ to $\mp$.

If the position indices would be instead 1 and 2 referring both to
outgoing  states, then the product representations would describe
states, where one would have

${\cal V}_t$: opposite helicities, symmetric,

${\cal A}_t$: opposite helicities, antisymmetric,

${\cal T}^{\pm}_t$: parallel helicities of two orientations.

The invariant out of two vectors (of any types) is built as
\beq g^{r,s} v_r v_s, \ \, r,s = 1,2,3,  (g^{r,s}) = {\rm diag} (1,1,-1)
\eeq
In this way three different symmetric four-parton helicity operators
\beqar
\left ( \matrix{A_1^* \cr A_1 \cr} \right )^T
\left ( \matrix{A_2^* \cr A_2 \cr} \right )^T
\hat O_{11^{\prime}, 2 2^{\prime } }
\left ( \matrix{A_{1^{\prime }}^* \cr A_{1^{\prime }}\cr} \right )
\left ( \matrix{A_{2^{\prime }}^* \cr A_{2^{\prime }} \cr} \right )
\eeqar
can be constructed with $\hat O_{1,1^{\prime}, 2 2^{\prime } }$
being replaced by the projectors
\beq
\hat {\cal V}_s = \hat {\cal V}^{1 1^{\prime }} \otimes \hat {\cal V}^{2
2^{\prime }},
\hat {\cal A}_s = \hat {\cal A}^{1 1^{\prime }} \otimes \hat {\cal A}^{2
2^{\prime }},
\hat {\cal T}_s = \hat {\cal T}^{+ 1 1^{\prime }} \otimes \hat {\cal
T}^{ - 2 2^{\prime }} +
 \hat {\cal T}^{ - 1 1^{\prime }} \otimes \hat {\cal T }^{ + 2
2^{\prime }}.
\eeq
The latter three operators correspond to definite $SU(1,1)$ helicity
states in $s$-channel ($1 1^{\prime } \rightarrow 2 2^{\prime} $). The
operators with definite helicity states in other channels can be written
in the same way. We quote the crossing relations between the channels
 $s$ ($1 1^{\prime } \rightarrow 2 2^{\prime} $),
 $u$ ($1 2^{\prime } \rightarrow 2 1^{\prime} $) and
 $t$ ($1^{\prime } 2^{\prime } \rightarrow 1 2 $) \cite{FK}
\beqar
\hat {\cal V}_s = \hat {\cal V}_u  = \hat {\cal T}_t + \hat {\cal V}_t +
\hat {\cal A}_t, \cr
\hat {\cal A}_s = - \hat {\cal A}_u  = \hat {\cal T}_t - \hat {\cal V}_t
- \hat {\cal A}_t, \cr
\hat {\cal T}_s = \hat {\cal T}_u  =  \hat {\cal V}_t -
\hat {\cal A}_t.
\eeqar

Obvously, the parallel helicity operators (\ref{ttgf}) are of the form ${\cal
T}_t$ and the antiparallel helicity operators (\ref{hgg}- \ref{hannih})
 decompose into ${\cal V}_t$ and ${\cal A}_t$.
 In particular all pure gluonic operators can be
written in the following $s$-channel form:
\beqar
\left ( \matrix{A_1^* \cr A_1 \cr} \right )^T
\left ( \matrix{A_2^* \cr A_2 \cr} \right )^T
\hat O_{1 1^{\prime}, 2 2^{\prime } } \
\partial
\left ( \matrix{A_{1^{\prime }}^* \cr A_{1^{\prime }}\cr} \right )
\partial
\left ( \matrix{A_{2^{\prime }}^* \cr A_{2^{\prime }} \cr} \right ),
\cr
\hat G_{1 1^{\prime} 2 2^{\prime }} =
(T^{a}_{1 1^{\prime}} \otimes T^{a}_{2 2^{\prime }})
\ \ \ \ \ \ \ \ \ \ \ \ \ \ \ \ \ \ \ \ \ \ \ \ \cr
\left \{-8  [ \tilde J_{1 1^{\prime }}^{(g)} + w_g^{(0)} \delta^{(2)} ]
\hat {\cal V}_s +
2 [\tilde J_{221} ] (\hat {\cal V}_s - \hat {\cal A}_s + 2 \hat {\cal T}_s
)  - 4 [\tilde J_{112} ] ( \hat {\cal V}_s - \hat {\cal A}_s )
\right \}
\eeqar
The operator
\beq
\hat {\cal V}_s - \hat {\cal A}_s + 2 \hat {\cal T}_s =\hat {\cal V}_t =
\sigma_1^{(1 1^{\prime} )} \otimes \sigma_1^{(2 2^{\prime }) }
[ I^{(1 1^{\prime })} \otimes I^{(2 2^{\prime } ) }  +
g^{r s} \sigma_r^{(1 1^{\prime} )} \otimes \sigma_s^{(2 2^{\prime })} ]
\eeq
has appeared already above in the double-logarithmic asymptotics
(\ref{dloghamiltonian}).

\subsection{Two-parton eigenstates}

Two-parton states on which the above operators act have the form
\beq
| \phi_{12} > = \sum_{P_i} \int dz_1 dz_2 \ \phi^{P_1 P_2} (z_1, z_2)
\ A^{(+)}_{P_1} (z_1) \ A^{(+)}_{P_2} (z_2) \ | 0 >.
\eeq
$P = (p, \lambda) $ labels the parton type $p$ (chirality and flavour)
and the helicity $\lambda$, i.e. $A_P = (A^*,  A, f^*, f,  \tilde f^* ...)$.
The action of the symmetries on the fields induces an action on $\phi$.
In particular the action of the one-dimensional conformal
transformations (\ref{confz}, \ref{conffields})
 on $\phi$ is generated by $ S^{a} = S_1^{a} +
S_2^{a}, a= \pm, 0$
\beq
S^-_1 = \partial_1, \ \ \ S^+_1 = z_1^2 \partial_1 + 2 s_{p1} z_1,
\ \ \  S^0_1  = z_1 \partial_1 + s_{p1},
\eeq
where $s_A = \frac{1}{2}, s_f = 1$.

Conformal symmetry is used to find the eigenstates and eigenvalues of
the above operators. Among the states with the same eigenvalues the ones
of lowest weight are distinguished,
\beqar
(S^-_1 + S^-_2) \ \phi^{P_1 P_2 }(z_1, z_2) = 0, \cr
(S^0_1 + S^0_2) \ \phi^{P_1 P_2 }(z_1, z_2) = (n + s_{p1} + s_{p2})
\  \phi^{P_1 P_2 }(z_1, z_2), \cr
 \phi^{P_1 P_2 }(z_1, z_2) \ = \ C_n^{P_1 P_2} \cdot (z_{12})^n.
\eeqar
Due to the helicity symmetry the eigenstates can be chosen to be
definite helicity states ${\cal T}^{\pm}, {\cal V}, {\cal A} $,
therefore
\beqar
C_{n, T^{\pm}}^{(p_1, \lambda_1) (p_2, \lambda_2)}  =
\tilde C_{n, T^{\pm}}^{p_1 p_2}  \ { 1 \pm \lambda_1 \over 2}
  \delta_{\lambda_1, \lambda_2},  \cr
C_{n, V}^{(p_1, \lambda_1) (p_2, \lambda_2)}  =
\tilde C_{n, V}^{p_1 p_2}  \ \delta_{\lambda_1, -\lambda_2},  \cr
C_{n, A}^{(p_1, \lambda_1) (p_2, \lambda_2)}  =
\tilde C_{n, A}^{p_1 p_2}  \ \ \lambda_1
  \delta_{\lambda_1, - \lambda_2}.
\eeqar
Clearly, the states ${\cal T}^{\pm}$ have the same eigenvalue.

If flavour or chirality do not coincide, $p_1 \not =  p_2$, then still
the diagonalization in that subspace has to be performed.

Comparing to the Regge limit we see that conformal symmetry holds in the
corresponding logarithmic approximation to both limits. In the Regge
case the operators depend on the transverse position  (impact
parameter) and
the conformal symmetry acts on the plane by M\"obius transformations
(based on $sl(2, C ) $). The operators of reggeon interactions allow for
factorization of the holomorphic and anti-holomorphic dependences on the
complex position variable. The leading reggeized gluons do not carry
helicity and there is no similarity regarding to the helicity symmetry.

\section{Discussion}
\setcounter{equation}{0}

We have treated the high energy Bjorken asymptotics of QCD amplitudes
choosing the non-standard viewpoint of the effective action concept. While
this does not add new results to what is well known about scale dependence
of composite operators or (ordinary and skewed) parton distributions it
allows to draw interesting parallels to the Regge asymptotics and provides 
a compact and uniform treatment of all exchange channels. 
A motivation is to rely on the observed parallels in improving the
understanding of the more involved Regge asymptotics.

The effective action for the Bjorken asymptotics is obtained from the QCD
action (\ref{action}) by separation of momentum modes. 
In the approximation (\ref{reduction},\ref{fieldasymp+},\ref{fieldasymp-})
it allows to calculate the interaction of the exchanged partons in the
tree approximation.

The results have been formulated in terms of Hamiltonian operators 
involving
light-ray kernels in a convenient Feynman parameter representation. 
In this operator form the symmetries of the one-dimensional 
effective parton interaction are manifest, providing a convenient basis for
treating the mutiparton exchange contributions of higher twist by symmetry
methods.
Contributions to the asymptotics are calculated by studying the
evolution induced by these Hamiltonians.

The eigenvalues are calculated easily in the $\alpha $ representation
(\ref{Jkernels}) reproducing the well known results for the sets of
anomalous dimensions in the exchange channels of parallel helicities
${\cal T}_t$ related to transversity asymmetry or chirality odd structure
functions like $h_1, F_3^{\gamma}$,
symmetric antiparallel helicity ${\cal V}_t$, related to unpolarized  
parton distributions, and antisymmetric antiparallel helicity ${\cal A}_t$,
related to helicity asymmetry and structure functions $g_1$. 
The parton states are directly related to the conventional local light-cone
operators composed out of the gluon field strength operator 
and the Dirac field operators with derivatives acting on
them. In the leading twist case the field strength enters with one index
contracted with $q^{\prime}$ and one being transverse and expessing the
polarization state, the chirality and polarization states of the quarks are
selected by the known $\gamma$ matrix projectors, among them a factor 
$\gamma_{\mu} q^{ \prime \mu} $, and the derivatives are all
longitudinal, $q^{\prime \mu} \partial_{\mu} $. The light-ray wave functions
are directly related to the derivative structure of the operator: The number
of derivatives acting on the $i$th operator is the power of $z_i$ in the
polynomial $\phi(z_i) $.

In the considered approximation the effective action and the resulting
Hamiltonians describe the twist-two (two-parton exchange) contributions
to the Bjorken asymptotics and moreover the part of higher twist (in the
power counting sense) contributions which results from the exchange of a
fixed number of partons equal to the twist. This corresponds to the
renormalization of quasi-parton operators \cite{BuFKL} composed of more
then two fields disregarding their mixing with other operators not of
the quasi partonic class.

There are contributions of t-channel intermediate states with  a number
of partons less than the twist and therefore the parton number may vary
in the iteration. However, since the twist is one and the same for all
intermediate states, those intermediate states with a smaller number 
of partons necessarily involve partons of non-leading type.
Non-leading exchanged partons  differ from the ones
considered so far and appear from improving the condition 
(\ref{reduction}, \ref{fieldasymp+}, \ref{fieldasymp-} ) as
mentioned in Section 2.3.  They are the analogon of the non-leading
reggeons. The contribution of the latter to the effective action of
Regge asymptotics has been studied recently \cite{eff}.
There is an important difference: In the Regge asymptotics the exponent
of $s$ of a definite reggeon exchange depends on the number of leading
reggeized gluons only by ${\cal O} (g^2) $ corrections. Therefore in the
interation one may have a variable number of those leading gluonic
reggeons; a variable reggeon number may appear without involving the
non-leading type of reggeons. In the Bjorken asymptotics the analogon of
the generalized leading log approximation is incomplete without
non-leading exchange partons.

The effective action for the Bjorken limit in the approximation
presented here is to be improved by including higher order effective
vertices and renormalization $Z$ factors and by terms describing the
non-leading partons and their interactions. The first improvement is to
reproduce in the next order in particular the two-loop anomalous
dimensions and evolution kernels well known in the standard approach.
The second improvement would be necessary for studying higher twist and
deserves further study.

 The multi-parton exchange 
(higher twist)   contribution with a fixed number leads to 
a multi-body quantum problem in one dimension
the hamiltonian of which is the sum of the above pair-interaction hamiltonians.
The solution of this stationary state quantum problem allows to construct
the corresponding higher twist contribution to the Bjorken limit amplitude.

This appears again in parallel to the Regge case.
In both cases some of the arising multi-body quantum systems turn out to be
completely integrable (without or with simplification of the gauge group
factors by taking the $N \rightarrow \infty $ limit). In these cases the
involved pair hamiltonians are particular operators out of complete sets of
integrals of motion which can be constructed from solutions of the
Yang-Baxter equation.

The pair interaction hamiltonians of the parallel helicity partons
(\ref{ttgf})  (without the gauge group factors) correspond to integrable
multi-body systems with nearest neighbour (chain) interactions
constructed from $sl(2)$ symmetric solutions of the Yang-Baxter equation. 
Interesting three-parton channels have been
 studied in  \cite{BDKM}. 

The parallel helicity operators altogether (disregarding the gauge group
factors) appear in the integrable homogeneous closed chain constructed out
of a  chiral ($\ell =  b $) representation of the
superconformal group $sl(2|1)$ \cite{DKK}.
The natural question arises whether integrable systems can be constructed
the hamiltonians of which coincide with the ones of the other helicity
configurations ${\cal V}_t, {\cal A}_t$.

The symmetries of the asymptotic effective interaction are derived from the
symmetries of the underlying theory. The impact of extended symmetries
(compared to QCD) on the asymptotic interactions is a question of interest,
since the answer may be helpful in treating multiple exchanges. With this
motivation we are going to consider the models of ${\cal N}= 1, 2, 4$
supersymmetric Yang-Mills theory in a forthcoming paper. The present
effective action formulation allows for a simple and uniform treatment of
the Bjorken asymptotics of those models.

\vspace{2cm}

{\large \sl Acknowledgements }

We are grateful to G.P.~Korchemsky, L.N.~Lipatov, A.N.~Manashov and
D.~Robaschik for useful discussions.

\section{Appendix A}
\setcounter{equation}{0}

The general definition of the $\bar \alpha$ integrals appearing
in the leading loop contributions are
\beqar
J_{n_1 n_2 ...n_N} (x_1, x_2, ..., x_N) =
\int_{-\infty }^{+\infty} {d \bar \alpha \over 2 \pi i}
 [ \bar \alpha x_1 +1 - i \epsilon]^{-n_1}
[ \bar \alpha x_2  +1 - i \epsilon]^{-n_2} \cr
... [ \bar \alpha x_N +1 - i \epsilon]^{-n_N}.
\label{alphaint}
\eeqar
$n_i$ are positive integers. The Feynman parameter representation is
\beqar
J_{n_1 n_2 ...}(x_1, x_2, ...) =
{\Gamma (\sum n_i -1) \over \Gamma(n_1) \Gamma(n_2) ...\Gamma(n_N)}
\int {\cal D}^{(N)} \alpha \ \alpha_1^{n_1 -1}  \alpha_2^{n_2 -1}
\cr ...\alpha_N^{n_N -1} \
\delta(\sum \alpha_i x_i )
\label{alphafeynman}
\eeqar
In the latter integral the varibles $\alpha_i$ 
range from 0 to 1 and  ${\cal D}^{(N)} \alpha = d\alpha_1 ...d\alpha_N
\delta((1 - \sum \alpha_i ) $.

By decomposition into simple fractions one can reduce the number of factors
in the denominator of (\ref{alphaint}). In this way one derives relations
like
\beqar
J_{1 2}(x_1, x_2) = { x_1 \over x_1 - x_2} J_{1 1} (x_1, x_2), \cr
J_{1 1 1}(x_1, x_2, x_3) = { x_2 \over x_{23} } J_{1 1} (x_1, x_2)
 - { x_3 \over x_{23} } J_{1 1} (x_1, x_3) , \cr
J_{1 1 2}(x_1, x_2, x_3) = {- x_1 x_2 \over x_{13} x_{23} } J_{1 1} (x_1, x_2)
 + \left ({ x_2 \over x_{23} } + {x_1 \over x_{13}} \right ) J_{1 1 1} (x_1,
x_2, x_3 ) , \cr
J_{2 2 1}(x_1, x_2, x_3) = { x_3^2 \over x_{13} x_{23} } J_{1 1 1} (x_1, x_2,
x_3) \cr
 - {x_1 x_2 \over x_{12}^2 } \left ({ x_2 \over x_{23} } + {x_1 \over x_{13}}
 \right ) J_{1 1 } (x_1,x_2 ).
\label{relations}
\eeqar
In the above calculations we encounter $\bar \alpha$ integrals with
three arguments ($x_1 = \beta_1, x_2 = - \beta_2, x_3 = \beta_{1 1^{\prime
}}$), two cases with two arguments ($x_1 = \beta_1, x_2 = \beta_{1 1^{\prime
}}$ and $x_1 = \beta_1, x_2 = - \beta_2 $) and with one argument, which is
just a $\delta$ function.

Under Fourier transformation
\beq {\cal F}[\Phi(\beta_1, \beta_2, \beta_{1^{\prime}}, \beta_{2^{\prime
}}) ] = \tilde \Phi(z_1, z_2,z_{1^{\prime }},z_{2^{\prime }})
\eeq
defined by
\beqar
\tilde \Phi(z_1, z_2,z_{1^{\prime }},z_{2^{\prime }})
= \int {d\beta_1 d\beta_2 d\beta_{1^{\prime }} d\beta_{2^{\prime }} \over
 (2 \pi)^2 }
\delta( \beta_1 + \beta_2 -  \beta_{1^{\prime}} - \beta_{2^{\prime }} ) \
\cr
\exp[ i(z_{1^{\prime }} \beta_{1^{\prime }} + z_{2^{\prime }} \beta_{2^{\prime }} +
 - z_1 \beta_1   - z_2 \beta_2  )] ]
\Phi(\beta_1, \beta_2, \beta_{1^{\prime}}, \beta_{2^{\prime }})
\eeqar
the expressions involving $J$ appearing in the interaction kernels transform
as
\beqar
{\cal F} [J_{n_1 n_2 n_3} (\beta_1, -\beta_2, \beta_{1 1^{\prime }}) ]
= {\Gamma(n_1+n_2+n_3-1) \over \Gamma (n_1) \Gamma(n_2) \Gamma(n_3) }
\int {\cal D}^{(3)} \alpha \ \cr
 \alpha_1^{n_1-1} \alpha_2^{n_2-1} \alpha_3^{n_3-1} \
\delta (z_{1 1^{\prime }} - \alpha_1 z_{12}) \
\delta (z_{2 2^{\prime }} + \alpha_2 z_{12}), \cr
{\cal F} [J_{1 1} (\beta_1, \beta_{1 1^{\prime }})  \varphi_1 (\beta_i) ]
= \int d \alpha \ \chi_1 ( \alpha )
\delta(z_{1 1^{\prime }} - \alpha z_{12}) \
\delta(z_{2 2^{\prime }} ), \cr
{\cal F} [J_{1 1} (\beta_1, -\beta_2 )  \varphi_0 (\beta_i) ]
=  \int d \alpha \  \ \chi_0 (\alpha )
\delta(z_{1 1^{\prime }} - \alpha z_{12}) \
\delta(z_{2 2^{\prime }} + (1-\alpha ) z_{12}), \cr
{\cal F} [\delta (\beta_1, -\beta_2 ) ]
=  \delta(z_{1 1^{\prime }} ) \ \delta(z_{2 2^{\prime }}).
\label{fourier}
\eeqar
$\beta $ dependent factors multiplying the functions $J_{11}$ result in
some additional functions in the intergrand of the Fourier transformed
Feyman parameter representations:
\beqar
\varphi_1 (\beta_i) = 1 \ \ \ \rightarrow \ \ \ \chi_1 (\alpha) = 1, \cr
\varphi_1 (\beta_i) = {\beta_1 \over \beta_{1^{\prime}} }
 \ \ \ \rightarrow \ \ \ \chi_1 (\alpha) = 1 - \alpha, \cr
\varphi_1 (\beta_i) = {\beta_1 \over \beta_{1 1^{\prime }} }
 \ \ \ \rightarrow \ \ \ \chi_1 (\alpha) =  - { 1 - \alpha \over \alpha }, \cr
\varphi_1 (\beta_i) = {\beta_1^2 \over \beta_{1^{\prime }} \beta_{1 1^{\prime }} }
 \ \ \ \rightarrow \ \ \ \chi_1 (\alpha) =  - { (1 - \alpha)^2 \over \alpha }, \cr
\varphi_0 (\beta_i) = 1 \ \ \ \rightarrow \ \ \ \chi_0 (\alpha) = 1, \cr
\varphi_0 (\beta_i) = {\beta_1 \over \beta_1 + \beta_2}
\ \ \ \rightarrow \ \ \ \chi_0 (\alpha) = 1 - \alpha, \cr
\varphi_0 (\beta_i) = {\beta_2 \over \beta_1 + \beta_2}
\ \ \ \rightarrow \ \ \ \chi_0 (\alpha) =  \alpha, \cr
\varphi_0 (\beta_i) = {\beta_1 \beta_2 \over (\beta_1 + \beta_2)^2 }
\ \ \ \rightarrow \ \ \ \chi_0 (\alpha) =  \alpha (1- \alpha), \cr
\eeqar

\section{Appendix B}
\setcounter{equation}{0}

In the case of antiparallel helicity gluons the calculation of the
effective two-parton interaction results instead of (\ref{vertext1} )
in \beqar
\left ( {\partial_{1^{\prime }}^2 \partial_{2^{\prime }}^2 +
\partial_1^2 \partial_2^2 \over (\partial_1 + \partial_{1^{\prime}})^2 }
\ \tilde J_{111} +
 { (\partial_{1^{\prime }} \partial_{2^{\prime }} +
\partial_1 \partial_2) \partial_1 \partial_2 \over
(\partial_1 + \partial_{1^{\prime}})^2 }
\ \tilde J_0 \right )
(A_1^* T^{a} A_{1^{\prime }} ) \ (A_2 T^{a} A_{2^{\prime }}^* )  \cr
 -  { ( \partial_{1} \partial_{1^{\prime }} +
\partial_2 \partial_{2^{\prime }}) \partial_1 \partial_2 \over
(\partial_1  + \partial_{2})^2 }
\ \ \tilde J_0 \ \
(A_1^* T^{a} A_{2} ) \ (A_{1^{\prime }} T^{a} A_{2^{\prime }}^* )  \cr
+ (\partial_1 + \partial_{1^{\prime}} )^2 \ \ \tilde J_{111} \ \
(A_1^* T^{a} A_{1^{\prime }}^* ) \ (A_2 T^{a} A_{2^{\prime }} )
\label{vertexa1}
\eeqar
Here the self-energy contributions are not included yet. We separate in
the first bracket a term equal to the result for the parallel helicity
case (\ref{vertext1}). Adding now the self-energy contribution
(\ref{disconnected}) results in the replacement of the first bracket by
\beq
2 \partial_{1^{\prime }} \partial_{2^{\prime }}   [\tilde J_{1 1^{\prime
}} + w_g^{(0)} \delta^{(2)} ] - (\partial_{1^{\prime }} -  \partial_1)
 (\partial_{2^{\prime }} -  \partial_2) \ \tilde J_{111} +
\partial_1 \partial_2   \tilde J_0 ,
\eeq
where now each term represents a regular operator.

The last two relations given in (\ref{relations}) for the $J$ integrals
Fourier transformed to light ray variables can be used now to substitute
$ (\partial_{1^{\prime }} -  \partial_1)
 (\partial_{2^{\prime }} -  \partial_2) \ \tilde J_{111} $ and
$ (\partial_{1^{\prime }} +  \partial_1)^2 \ \tilde J_{111} $
by $\tilde J_{112}$ and $\tilde J_{221}$ plus remainders involving
$\tilde J_0$. This results in
\beqar
\left (2 [\tilde J_{1 1^{\prime }}+  w_g^{(0)} \delta^{(2)} ]
+ \tilde J_{221} -2 \tilde J_{112}  \right  ) \
(A_1^* T^{a} \partial A_{1^{\prime }} ) \ (A_2 T^{a} \partial A_{2^{\prime }}^* )  \cr
+ \tilde J_{221} \ \
(A_1^* T^{a} \partial A_{1^{\prime }}^* ) \ (A_2 T^{a} \partial A_{2^{\prime }} )  \cr
 +
 {\partial_1 \partial_2 \over
(\partial_1  + \partial_{2})^2 }
\tilde J_0 \ \  \{ -  (\partial_{1} \partial_{2^{\prime }} +
\partial_2 \partial_{1^{\prime }})
(A_1^* T^{a} A_{1^{\prime }}^* ) \ (A_2 T^{a} A_{2^{\prime }} ) \cr
 +  (\partial_{1} \partial_{1^{\prime }} +
\partial_2 \partial_{2^{\prime }})
[ (A_1^* T^{a} A_{1^{\prime }} ) \ (A_2 T^{a} A_{2^{\prime }}^* )
- (A_1^* T^{a} A_{2} ) \ (A_{1^{\prime }} T^{a} A_{2^{\prime }}^* ) ]
   \} .
\eeqar
Due to the commutation relation of the generators $T^{a} $ we have
for the gauge group brackets (\ref{brackets})
\beq
 (A_1^* T^{a} A_{1^{\prime }} ) \ (A_2 T^{a} A_{2^{\prime }}^* )
- (A_1^* T^{a} A_{2} ) \ (A_{1^{\prime }} T^{a} A_{2^{\prime }}^* )
=
(A_1^* T^{a} A_{2^{\prime }}^* ) \ (A_2 T^{a} A_{1^{\prime }} ).
\eeq
Thus the last  term in the brackets multiplying $\tilde J_0 $  in (9.3)
is equal
to the first  one up to the sign and the exchange of $1^{\prime }$ and
$2^{\prime }$.
We remember that the above expressions are understood as integrated over
the light ray positions $ 1, 2, 1^{\prime}, 2^{\prime}$.  Therefore the
exchange of $1^{\prime }$ and $2^{\prime }$ is just a substitution of
integration variables. We conclude that the contribution involving
$\tilde J_0$ cancels and arrive at the result (\ref{hgg}) up to
normalization.

The other cases of anti-parallel helicities are simpler. For the
quark-gluon and the annihilation-type interactions the following
relation is used to rearrange derivative operators,
\beqar
(\partial_2 + \partial_{2^{\prime} } ) \tilde J_{111} =  -2  \
\partial_{1^{\prime } } \tilde J_{211} + (\partial_{1^{\prime} }  +
\partial_{2^{\prime } } \tilde J_0^{(g)} \cr
= \  2 \ \partial_{2^{\prime } } \tilde J_{211} - (\partial_{1^{\prime} } +
\partial_{2^{\prime }} \tilde J_0^{(g)}.
\eeqar
We present typical calculation for transforming an extra derivative
acting on a kernel like $\tilde J_{111} $ into a factor $z_{12}^{-1} $
(transformation of (\ref{hannih}) to (\ref{annihc}) ),
\beqar
\partial_{2^{\prime }} \int {\cal D}^{(3)} \alpha \ \varphi (\alpha_1,
\alpha_2) \ \delta (z_{1 1^{\prime }}  - \alpha_1 z_{12} )
\ \delta (z_{2 2^{\prime }}  + \alpha_2 z_{12} ) \cr
= - z_{12}^{-1} \ \int
 {\cal D}^{(3)} \alpha \  \varphi (\alpha_1, \alpha_2) \
 \partial_{\alpha_2}
[\delta (z_{1 1^{\prime } }
 - \alpha_1 z_{12} ) \ \delta (z_{2 2^{\prime } } + \alpha_2 z_{12} ) ] \cr
=  z_{12}^{-1} \ \int
 {\cal D}^{(3)} \alpha \ \partial_{\alpha_2} \varphi (\alpha_1,
\alpha_2) \
\delta (z_{1 1^{\prime
} } - \alpha_1 z_{12} ) \ \delta (z_{2 2^{\prime } } + \alpha_2 z_{12} )
\cr
-  z_{12}^{-1} \ \int
 d \alpha_1 \  \varphi (\alpha_1,
 0) \
\delta (z_{1 1^{\prime
}  } - \alpha_1 z_{12} ) \ \delta (z_{2 2^{\prime } }  ) \cr
+
  z_{12}^{-1} \ \int
d \alpha_1 \  \varphi (\alpha_1,
1- \alpha_1) \
\delta (z_{1 1^{\prime
} } - \alpha_1 z_{12} ) \ \delta (z_{2 2^{\prime } } + (1-\alpha_1)  z_{12}
).
\eeqar
The simpler case with a kernel of the type $\tilde J_0 $ looks as
follows,
\beqar
(\partial_{1^{\prime } } + \partial_{2^{\prime }})  \ \int
d \alpha \  \alpha (1-\alpha) \chi (\alpha)
\delta (z_{1 1^{\prime
} } - \alpha z_{12} ) \ \delta (z_{2 2^{\prime } } + (1-\alpha)  z_{12}
) \cr
=
- z_{12}^{-1}   \ \int
d \alpha \  \partial_{\alpha} [ \alpha (1-\alpha) \chi (\alpha) ]
\delta (z_{1 1^{\prime
} } - \alpha z_{12} ) \ \delta (z_{2 2^{\prime } } + (1-\alpha )  z_{12}).
\eeqar

In order to arrive at (\ref{annihc}) a cancellation between $ s,t$ 
and $u$ channel type terms multiplying $ \tilde J_0^{(g)} $ 
analogous to  (9.4) has to be done.


\begin{thebibliography}{99}

\bibitem{JCPS} J.C. Collins. {\it Renormalization},  Cambridge Univ.
Press 1984 \\\
 M. E. Peskin and D. V. Schroeder, {\it An Introduction to Quantum
Field Theory}, Addison-Wesley Publ., 1995


\bibitem{GStBLKod} J.C. Collins, D.E. Soper and G. Sterman, {\it
Factorization and hard processes in QCD },
              in {\it Perturbative QCD },
ed. A.H. Mueller, World Scientific, Singapore,  1989, p.1 ;\\
 S.J. Brodsky and G.P. Lepage, in {\it Perturbative QCD},
 ed. A.H. Mueller, World Scientific 1989, p.93; \\
J. Kodaira, Progr. Theor. Phys. Suppl. 120 (1995) 37.


\bibitem{twist} A. Ali, V.M. Braun, G. Miller, Phys. Lett. B266 (1991)
117;\\
Ya. Ya. Balitsky, V.M. Braun, Y. Koike, T. Tanaka, Phys. Rev. Lett. 77
(1996) 3078;\\
P. Ball, V. Braun, Y. Koike, T. Tanaka, Nucl. Phys. B529(1998) 323.


\bibitem{SPD}  D. M\"uller, D. Robaschik, B. Geyer, F.M. Dittes, J.
Horejsi, Fortschr. Physik 42 (1994) 101; \\
A.V. Radyushkin, Phys. Lett. B 385 (1996) 333; \\
X. Ji, Phys. Rev. D55 (1997) 7714;\\
J.C. Collins, L. Frankfurt, M. Strikman, Phys. Rev. D56 (1997) 2982.



\bibitem{DGLAP}
V.G. Gribov and L.N. Lipatov, Sov. J. Nucl. Phys. 15(1972)438 \\
L.N. Lipatov,  Yad. Fiz. 20(1974)532     \\
G.~Altarelli and G.~Parisi, Nucl. Phys. B126(1977)298 \\
Yu.L. Dokshitzer, ZhETF 71(1977)1216

\bibitem{ERBL} V.L. Chernyak and A.R. Zhitnitsky, JETP Lett 25 (1977) 510;
\\
 A.V. Efremov, A.V. Radyushkin, Theor. Math. Phys. 42 (1980)
  97; Phys. Lett. B94 (1980) 245.
\newline
S.J. Brodsky, G.P. Lepage, Phys. Lett B87 (1979) 359; Phys. Rev. D22 (1980)
2157.


\bibitem{GR} B. Geyer, D. Robaschik, M. Bordag, J. Horejsi, Z. Phys. C26
(1985) 591; \\
T. Braunschweig, B. Geyer, J. Horejsi, D. Robaschik, Z. Phys. C33 (1987)
275, \\
F.M. Dittes, B. Geyer, D. M\"uller, D. Robaschik, J. Horejsi, Phys. Lett.
B209(1988) 325.



\bibitem{BFKL} L.N. Lipatov, Sov.J.Nucl.Phys. 23(1976)338          \\
               V.S. Fadin, E.A. Kuraev and L.N. Lipatov, \PL 60B(1975)50;
Sov.Phys. JETP 44(1976)443; {\it ibid} 45(1977)199 \\
                Y.Y. Balitski and L.N. Lipatov, Sov.J.Nucl.Phys. 28(1978)882



\bibitem{Lev89} L.N. Lipatov, {\it Pomeron in QCD}, in {\it Perturbative
QCD}, A.H. Mueller ed. , World Scientific 1989, p. 411.

\bibitem{Levrep} L.N. Lipatov, Phys. Reports 286 (1997) 131.



\bibitem{eff} L.N. Lipatov, \NP B365(1991)614; \\
 R. Kirschner, L.N. Lipatov and L. Szymanowski, \NP
     B452(1994)579; \PR D51(1995)838; \\
L.N. Lipatov, \NP B452(1995)369; \\
R. Kirschner and L. Szymanowski, Phys. Rev. D52(1995)2333;
 \PL B419 (1998) 348; Phys. Rev. D58(1998) 014004.




\bibitem{BuFKL} A.P. Bukhvostov, E.A. Kuraev and  L.N. Lipatov,
ZhETF 87 (1984) 37; \\
A.P. Bukhvostov, G.V. Frolov, E.A. Kuraev and L.N. Lipatov,
Nucl. Phys. B258 (1985) 601.





\bibitem{BB}  Ya. Ya. Balitsky and V.M. Braun, Nucl. Phys. B311 (1988/89)
541.

\bibitem{Makeenko}  Yu. M. Makeenko, Sov. J. Nucl. Phys. 33 (1981) 440; \\
Th. Ohrndorf, Nucl. Phys. B198 (1982) 26.

\bibitem{Muller} D. M\"uller, Phys. Rev. D49 (1994) 2525; D51 (1995) 3855;
D58 (1998) 054005; \\
A.V. Belitsky and D. M\"uller, Nucl. Phys. B537 (1999) 397.


\bibitem{BGR}  J. Bl\"umlein, B. Geyer and D. Robaschik, Phys. Lett
B406(1997) 161; Nucl. Phys. B560 (1999) 283; and \\
J. Bl\"umlein and D. Robaschik, Nucl. Phys. B 581 (2000) 449. 


\bibitem{BFM}  A.V. Belitsky, A. Freund and D. M\"uller, Nucl. Phys. B574
(2000) 347.


\bibitem{LevPadua} L. N. Lipatov, Padua preprint DFPD/93/TH/70, hep-th/9311037
(unpublished);
  JETP Lett. 59(1994)571.




\bibitem{BDKM} V.M. Braun, S.E. Derkachov, A.N. Manashov, Phys. Rev. Lett 81
(1998) 2020; \\
 V.M. Braun, S.E. Derkachov, G.P. Korchemsky, A.N. Manashov,
Nucl. Phys. B553 (1999) 355; \\
S.~E.~Derkachov, G.~P.~Korchemsky and A.~N.~Manashov,
Nucl.\ Phys.\  { B566} (2000) 203; \\ 
V.~M.~Braun, G.~P.~Korchemsky and A.~N.~Manashov, Nucl. Phys. B597 (2001)
370.


\bibitem{Belitsky}  A.V. Belitsky, Phys. Lett B453 (1999) 59.




\bibitem{BKP} J. Bartels, Nucl. Phys. B175(1980)365  \\
   J. Kwiecinski and M. Praszalowicz, Phys. Lett. 94B(1980)413




\bibitem{FL} V.S.~Fadin and L.N.~Lipatov,
 Nucl. Phys. B406(1993)259;
\NP B447 (1996) 767; Phys. Lett. B429 (1998) 127.







\bibitem{doublelog} R. Kirschner and L.N. Lipatov, Nucl. Phys. B213 (1993)
122; \\
J. Bartels, B.I. Ermolaev and M.G. Ryskin, Zeitschr.f.Phys. C70 (1996) 273;
Zeitschr. f. Phys. C72 (1996) 627; \\
R. Kirschner, L. Mankiewicz, A. Sch\"afer, L. Szymanowski, Zeitschr.f. Phys.
C74 (1997) 501;\\
B.I. Ermolaev, R. Kirschner and L. Szymanowski, Eur.Phys. J. C7 (1999) 65.

\bibitem{FK} M. Fippel and R. Kirschner, J. Phys. G17(1991)421.


\bibitem{DKK} S. Derkachov, D. Karakhanyan and R. Kirschner, 
 Nucl. Phys. B 583 [FS] (2000) 691; and Leipzig preprint NTZ 1/2001,
nlin.SI/0102024.



\end{thebibliography}
\end{document}